\documentclass[pra,twocolumn,floatfix]{revtex4}
\usepackage{graphicx}
\usepackage{color}
\usepackage{amsmath}
\usepackage{amssymb}
\DeclareUnicodeCharacter{2009}{\,}
  
\begin{document}

\title{Phases of a rotating Bose-Einstein condensate trapped in a hard-wall potential}
\author{H\'el\`ene Perrin$^{1,2}$ and G. M. Kavoulakis$^{3,4}$}
\affiliation{$^1$Universit\'e Sorbonne Paris Nord, Laboratoire de Physique des Lasers,
99 av. J.-B. Cl\'ement, F-93430 Villetaneuse, France
\\
$^2$CNRS, UMR 7538, F-93430, Villetaneuse, France
\\
$^3$Department of Mechanical Engineering, Hellenic Mediterranean University, 
P.O. Box 1939, GR-71004, Heraklion, Greece
\\
$^4$HMU Research Center, Institute of Emerging Technologies, GR-71004, Heraklion, Greece}
\date{\today}

\begin{abstract}

We investigate the rotational properties of a zero-temperature Bose-Einstein 
condensed gas, which is confined in a hard-wall potential. For weak interactions the gas 
carries its angular momentum via vortices of multiple quantization. As the coupling 
increases and for sufficiently slow rotational frequencies of the trap, the 
multiply-quantized vortex states become unstable against a mixed state. For stronger 
couplings, as the angular frequency of the trap increases, there is a transition from 
a vortex lattice, to a vortex lattice with a hole at the trap center, and finally to 
a giant vortex. For weak effective attractive interactions we still have vortices of 
multiple quantization. These states become unstable towards states which break the axial 
symmetry of the problem and resemble center-of-mass excitation, as the coupling becomes 
more negative. Finally, for even more attractive forces, the system becomes unstable 
against collapse.

\end{abstract}

\maketitle

\section{Introduction}

The rotational properties of superfluids have been the subject of numerous 
experimental and theoretical studies over the last decades \cite{AJL}. In a  
seminal experimental study, the formation of vortex states was recorded in 
liquid Helium IV \cite{Helium}. In this experiment, liquid Helium IV was 
confined in a bucket, and as the angular velocity of the container increased, 
the number of vortices increased, as well.

More recently, vortex states have also been created and observed in atomic gases, 
see, e.g., Refs.\,\cite{b0,b00,b1,b2,b3,b4,b5,b6}. In the initial experiments the 
confining potential was harmonic, see, e.g., Ref.\,\cite{Fetter2009}. More recently, 
however, it has become possible to design other forms of confining potentials, 
including anharmonic \cite{Bretin2004,Perrin,Perrin2}, toroidal/annular \cite{c1,
c2,c3, c4,c5,c6,c7,c8,c9,c10}, and hard-wall potentials \cite{Gaunt2013, Dalnat, 
Navon2021}. 

Motivated by the experiment on Helium IV \cite{Helium}, one question that arises 
naturally is how an atomic Bose-Einstein condensate which is confined in a bucket 
responds to rotation, given also that such a potential has been designed experimentally 
\cite{Gaunt2013, Dalnat, Navon2021}. The problem of a three-dimensional Bose gas 
rotating in a hard-wall cylinder at zero temperature has been addressed (variationally) 
in the Thomas-Fermi regime in Ref.\,\cite{e4}, which has identified three phases: a vortex 
lattice, a vortex lattice with a central hole, and a giant vortex, see also Ref.\,\cite{NR}. 
One may wonder if this still holds away from the Thomas-Fermi regime, when interactions are 
weaker, or even attractive. According to the results of the present study, a substantially 
richer set of phases emerges at arbitrary interaction. Reference \cite{e11} has 
examined this problem for weak and repulsive interactions. Also a relevant recent 
paper is that of Ref.\,\cite{Machado}, where the stability and dynamics of vortices 
in a flat potential was examined, using beyond-mean-field approaches.

In the present study we consider zero temperature and assume a very strong harmonic 
confining potential along the axis of rotation, with an oscillator length $a_z$ and 
as a result the problem becomes quasi-two-dimensional. Furthermore, we choose to vary 
two parameters, namely the angular velocity of the container $\Omega$, as well as the 
coupling constant $g$ (in the experiments on atomic systems both $\Omega$ and $g$ are 
tunable). In particular, in atomic gases the scattering length $a_{\rm sc}$ for s-wave 
elastic atom-atom collisions may be tuned via the so-called Feshbach resonances, see, 
e.g., \cite{Fesh} (we should recall that $g \propto a_{\rm sc}$ provided that 
$|a_{\rm sc}| \ll a_z$). In liquid Helium IV, on the other hand, tuning the 
coupling constant is not possible. 

In the case of hard-wall confinement, the single-particle energy $\epsilon_m$ has 
a positive curvature as a function of the single-particle angular momentum $m \hbar$. 
As a result, the single-particle energy in the rotating-frame $\epsilon_m^{\rm rot} 
= \epsilon_m - m \hbar \Omega$ has a unique minimum at a specific value $m_0$, for 
a fixed value of $\Omega$. Therefore, for zero, or sufficiently weak -- repulsive 
or attractive -- interatomic interactions, the gas occupies the lowest single-particle 
state of the hard-wall potential, with angular momentum $m_0 \hbar$, which corresponds 
to a multiply-quantized vortex state. Furthermore, as the angular frequency of the trap 
$\Omega$ increases, a transition between the single-particle orbitals with a winding 
number $m_0$ and $m_0 + 1$ takes place, discontinuously \cite{e11}. 

In addition, for a fixed value of $\Omega$, as the interaction strength $g$ 
increases, with $g > 0$, a multiply-quantized vortex state becomes energetically 
unfavourable -- see, e.g., Refs.\,\cite{Kavoulakis2000, e11} -- resulting in a 
continuous phase transition. In this phase transition, a multiply-quantized 
vortex state with circulation $m_0$ becomes unstable against a ``mixed" state, 
i.e., a linear superposition of three single-particle orbitals, with a winding 
number $0, m_0$, and $2 m_0$ \cite{e11} (due to conservation of angular
momentum in the collisions). 

What we have mentioned so far refers to weak interactions. In addition to the 
two phases that we just described, more phases emerge at stronger coupling
\cite{e4}. As the coupling $g$ increases, the system enters the Thomas-Fermi 
limit, where the healing length $\xi$ becomes much smaller than the size of the 
container $R$, $\xi/R \approx 1/\sqrt{8 g}$. In this regime, as the angular velocity 
of the bucket increases, we have the formation of a vortex lattice. For a fixed value 
of $g$, as $\Omega$ increases, the centrifugal potential creates a hole in the 
middle of the vortex lattice, i.e., we have a phase with a vortex lattice with 
a hole. For even larger values of $\Omega$, the lattice disappears and the system 
turns to a giant vortex. This phase resembles the state of multiple quantization 
(which we described above, for zero or weak interactions), however is distinct, 
as we argue below. The state with a vortex lattice, the one with a 
lattice with a hole at the middle of the  cloud and the giant vortex have also been 
found in the case of anharmonic trapping potentials, see, e.g., 
Refs.\,\cite{e1,e2,e3,e5,e6,e7,e8,e9,e10}.

For an effective attractive interatomic interaction, $g < 0$, there are additional 
phases beyond those described above. In this case, while a homogeneous system 
is unstable against collapse, the presence of the confining potential may stabilize 
it, while rotation enhances the stability even more. As a result, for sufficiently 
small values of $|g|$ the energy is still dominated by the single-particle energy 
and we still have vortices of multiple quantization. As $|g|$ increases, it becomes 
energetically favourable for the gas to break the axial symmetry of the problem 
and form a localized blob. Thus, this phase resembles center-of-mass excitation. 
Finally, for even larger values of $|g|$ the system collapses. 

To summarize, one may identify seven distinct phases. The first one is that of multiple
quantization, the second is a mixed phase, the third is a vortex lattice, the fourth
is a vortex lattice with a hole, the fifth is a giant vortex, the sixth is a state of 
center-of-mass-like excitation and finally the seventh is the unstable/collapsed phase. 

In what follows, in Sec.\,II we introduce our theoretical model. In Sec.\,III, 
we start with the case of a zero or weak effective (repulsive, or attractive) 
interatomic interaction and present numerical and analytic results for the phase 
diagram. In the same section we then turn to the Thomas-Fermi limit of strong 
repulsive interactions. In Sec.\,IV we examine the case of an effective 
attractive interaction. There, in addition to the phase of multiply quantization 
(which is present also for weak repulsive interactions), we discuss the 
center-of-mass-like excitation, as well as the collapsed phase. In Sec.\,V we 
comment on the experimental relevance of our results and finally in Sec.\,VI we 
summarize our results.
  
\section{Model}

Our Hamiltonian includes the kinetic energy, the external potential, and the
interaction energy, which we assume to have the usual contact form, 
\begin{equation}
  {\hat H} = \sum_i \left( - \frac {\hbar^2} {2 M} {\nabla_i^2} + V({\bf r}_i) \right) 
  + \frac 1 2 {U_0} \sum_{i \neq j} \delta({\bf r}_i - {\bf r}_j).
\end{equation}
Here $M$ is the atom mass and $U_0$ is the matrix element for elastic atom-atom collisions, 
with $U_0 = 4 \pi \hbar^2 a_{\rm sc}/M$, where $a_{\rm sc}$ is the scattering length 
for elastic s-wave atom-atom collisions. 

We assume that the trapping potential $V({\bf r})$ has the form $V({\bf r}) = 
V_{\perp}(\rho, \theta) + V_z(z)$, where $\rho, \theta$, and $z$ are the usual 
cylindrical coordinates. We consider a very tight harmonic potential along the 
$z$ axis, which is the axis of rotation, $V_z(z) = M \omega_z^2 z^2/2$, with a frequency 
$\omega_z$. Because the trapping potential is considered to be very tight, 
the motion of the atoms along this direction is frozen and the problem becomes 
quasi-two-dimensional. Thus, the order parameter $\Phi(\rho, \theta, z)$ takes the 
product form,
\begin{equation}
  \Phi(\rho, \theta, z) = \phi_0(z) \, \Psi(\rho, \theta),
  \label{orpa}
\end{equation}
where $\phi_0(z) = e^{-z^2/(2 a_z^2)}/\sqrt[4]{\pi a_z^2}$ is the lowest-energy state of 
the harmonic potential along the $z$ axis, with $a_z = \sqrt{\hbar/(M \omega_z)}$ being
the oscillator length. The energy functional of the three-dimensional problem is 
\begin{eqnarray}
  {\cal E}(\Phi, \Phi^*) = \frac {\hbar^2} {2 M} \int |\nabla \Phi|^2 \, d^3 r 
  + \int |\Phi|^2 V_{\perp}(\rho) \, d^3 r +
  \nonumber \\
  + U_0 \int |\Phi|^4 \, d^3 r 
  - \mu_{\rm 3D} \int |\Phi|^2 \, d^3 r,
  \label{functional00}
\end{eqnarray}
where $\mu_{\rm 3D}$ is the chemical potential that corresponds to the initial, three 
dimensional problem. With the assumption of Eq.\,(\ref{orpa}) takes the form, 
\begin{eqnarray}
  {\cal E}(\Psi, \Psi^*) = \frac {\hbar^2} {2 M} \int |\nabla \Psi|^2 \, d^2 \rho 
  + \int |\Psi|^2 V_{\perp}(\rho) \, d^2 \rho +
  \nonumber \\
  + 2 \pi g \frac {\hbar^2} M \int |\Psi|^4 \, d^2 \rho - \mu \int |\Psi|^2 \, d^2 \rho,
  \label{functional0}
\end{eqnarray}
after one integrates over the $z$ direction. In the above equations $\mu$ is equal to 
$\mu_{\rm 3D} - \hbar \omega_z/2$. Also, $g = (1/\sqrt{2 \pi}) N a_{\rm sc}/a_z$, is our 
dimensionless coupling constant, where $N$ is the atom number. The condition for 
Eq.\,(\ref{orpa}) to be valid is $|\mu| \ll \hbar \omega_z$.

Regarding the trapping potential on the plane, $V_{\perp}(\rho, \theta)$, we consider 
a hard-wall potential, with some radius $R$,
\begin{equation}
V_{\perp}(\rho) = 
\begin{cases} 
      0 & \text{if} \,\, \rho < R \\
      \infty & \text{if} \,\, \rho \ge R. 
\end{cases}
\label{pot2}
\end{equation}
The form of $V_{\perp}(\rho)$, implies that the order parameter vanishes for $\rho=R$.

To get some insight, the ratio between the interaction energy of a 
non-rotating and homogeneous gas, $2 g \hbar^2/(M R^2)$, and the kinetic energy, 
$\hbar^2/(2 M R^2)$, is equal to $4 g$. The limit $|g| \ll 1$ corresponds to weak 
interactions, where the interaction energy is much smaller than the kinetic energy, 
while the opposite limit $g \gg 1$, corresponds to the Thomas-Fermi regime.

Furthermore, the condition for quasi-two-dimensional motion, $|\mu| \ll \hbar \omega_z$ 
(for zero temperature that we consider here) may be written as $N |a_{\rm sc}| a_z/R^2
\ll 1$, or $|g| \ll (R/a_z)^2$. An additional condition that has to be satisfied is 
$|a_{\rm sc}| \ll a_z$, in order for the atom-atom scattering to be treated as 
three-dimensional \cite{PHS, PS, OPL}. Combining the last two conditions one gets that 
the diluteness parameter $n |a_{\rm sc}|^3 \ll 1$, since $n |a_{\rm sc}|^3 \approx
(N |a_{\rm sc}| a_z/R^2) (a_{\rm sc}/a_z)^2$. As a result, the use of the mean-field 
approximation that we use is justified. We stress that the inequality $1 \ll g \ll 
(R/a_z)^2$ (for repulsive interactions) may easily be satisfied. In other words, the 
system may be in the Thomas-Fermi limit and still the mean-field approximation be valid.

We use some length scale $R_0$ as the unit of length and define $\Omega_0 = \hbar /(M R_0^2)$. 
In the numerical calculations we choose $R/R_0 = 3$. In the rest of the paper we set $\hbar = 
M = R_0 = 1$.
 
\begin{figure}
\includegraphics[width=9cm,height=7cm,angle=-0]{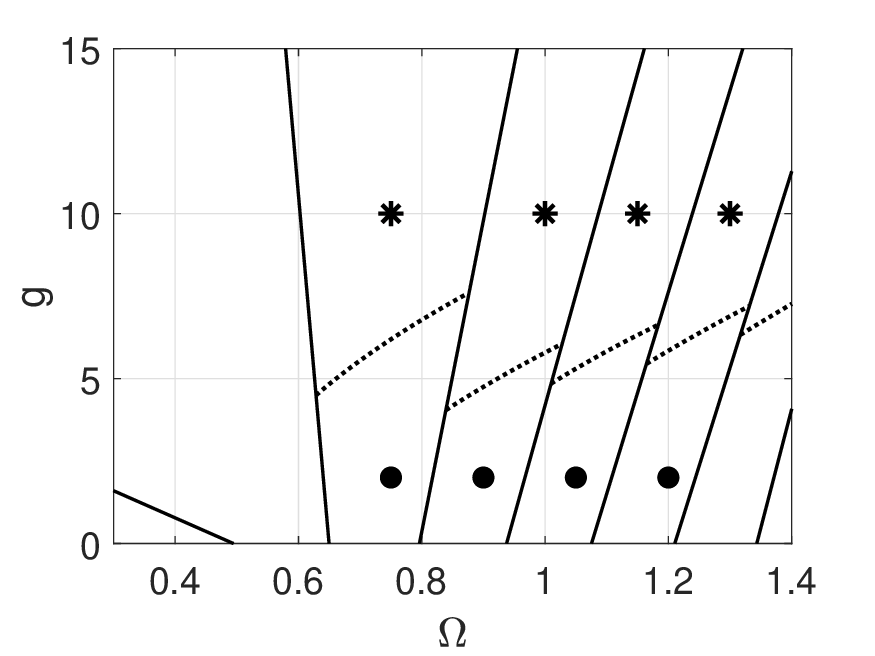}
\caption{Phase diagram, where on the $x$ axis is the rotational frequency of the trap 
$\Omega$ (in units of $\Omega_0 = \hbar/(M R_0^2)$) and on the $y$ axis is the 
(dimensionless) coupling $g$. Here we have a hard-wall trapping potential with $R/R_0 
= 3$. The solid lines, which are given by Eq.\,(\ref{neqq}), give the discontinuous 
transitions, while the dotted curves give the continuous ones. The dotted curves 
separate the phase diagram into two parts. In the lower part the order parameter 
corresponds to vortices of multiple quantization. In the upper part, the order 
parameter corresponds to a mixed state, see Eq.\,(\ref{mixed}). In the region of 
multiple quantization, each time one crosses a solid line, the circulation increases 
by one unit, starting from zero. The circles ``$\bullet$" correspond to the data 
shown in Fig.\,2, while the asterisks ``${\bf \ast}$" correspond to the data shown in 
Fig.\,4.}
\end{figure}

\begin{figure}
\includegraphics[width=4.1cm,height=3.27cm,angle=-0]{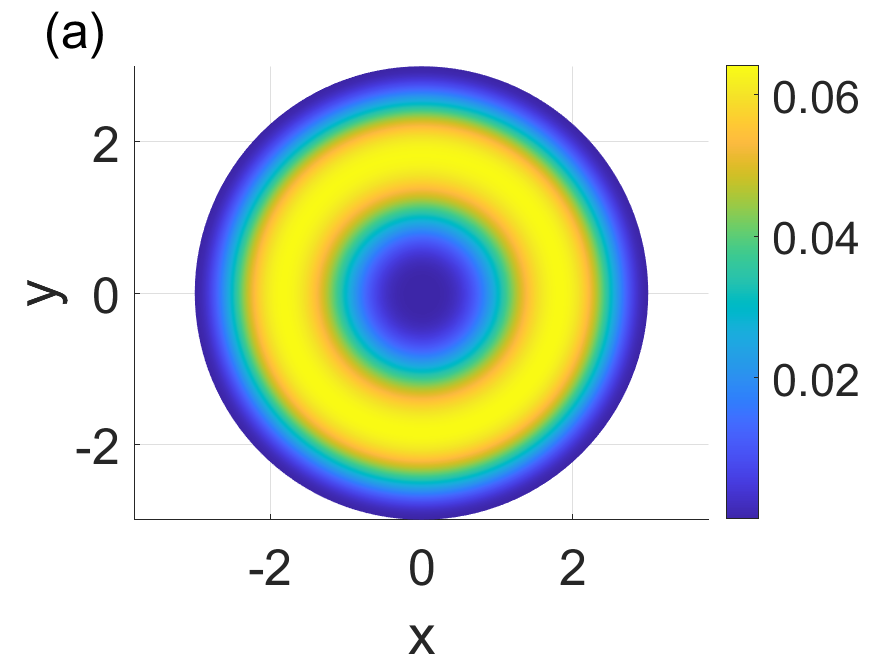}
\includegraphics[width=4.1cm,height=3.27cm,angle=-0]{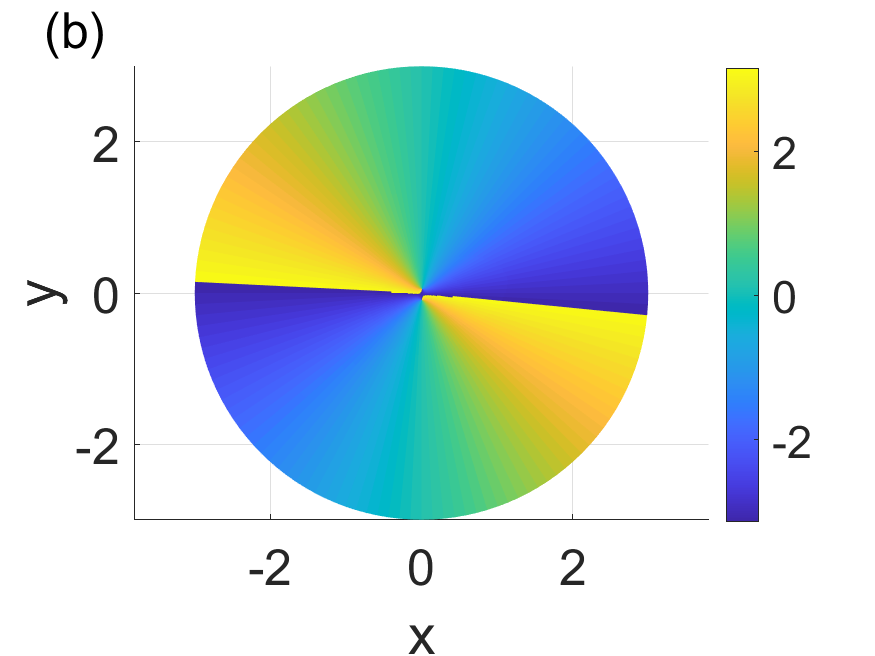}
\includegraphics[width=4.1cm,height=3.27cm,angle=-0]{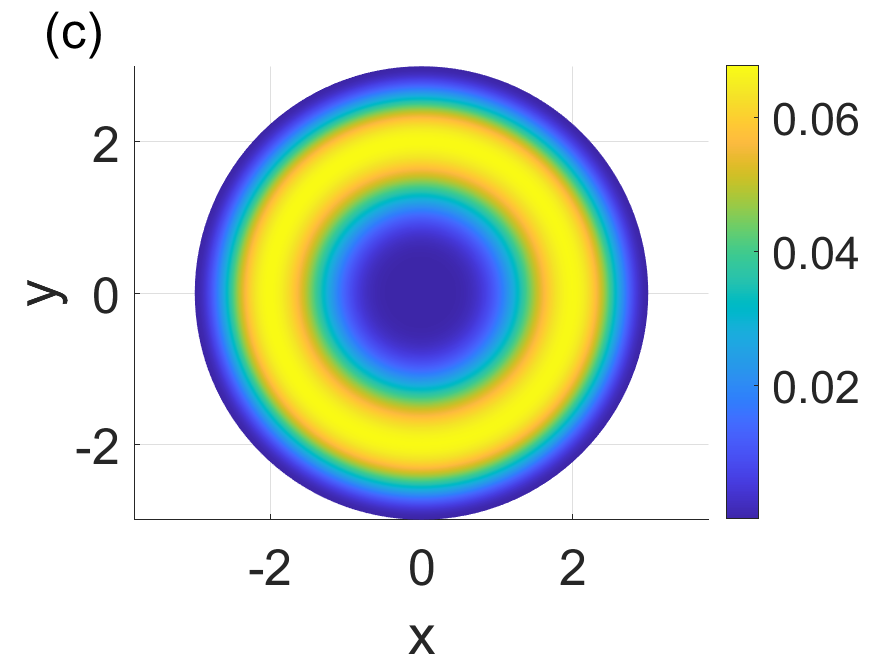}
\includegraphics[width=4.1cm,height=3.27cm,angle=-0]{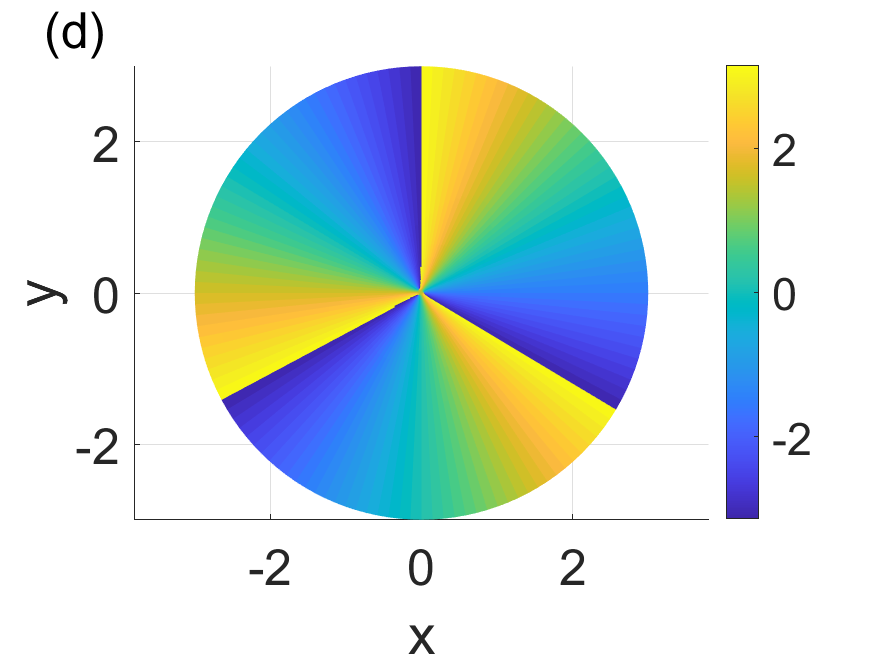}
\includegraphics[width=4.1cm,height=3.27cm,angle=-0]{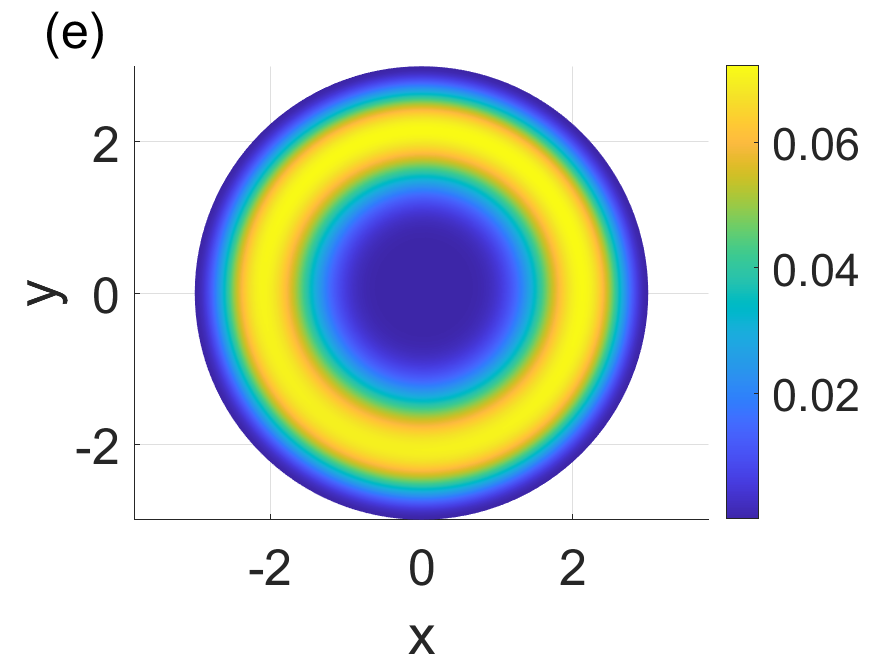}
\includegraphics[width=4.1cm,height=3.27cm,angle=-0]{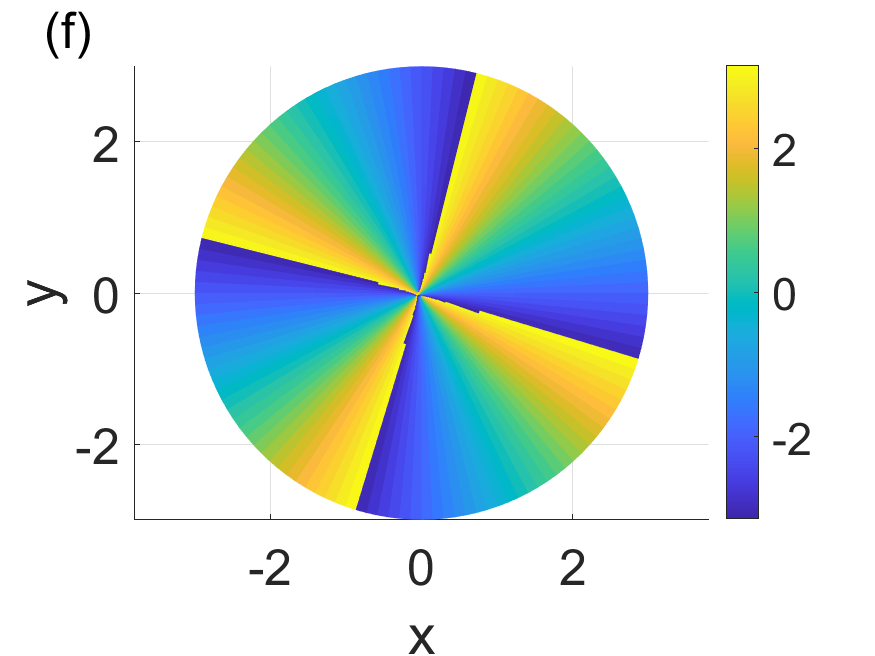}
\includegraphics[width=4.1cm,height=3.27cm,angle=-0]{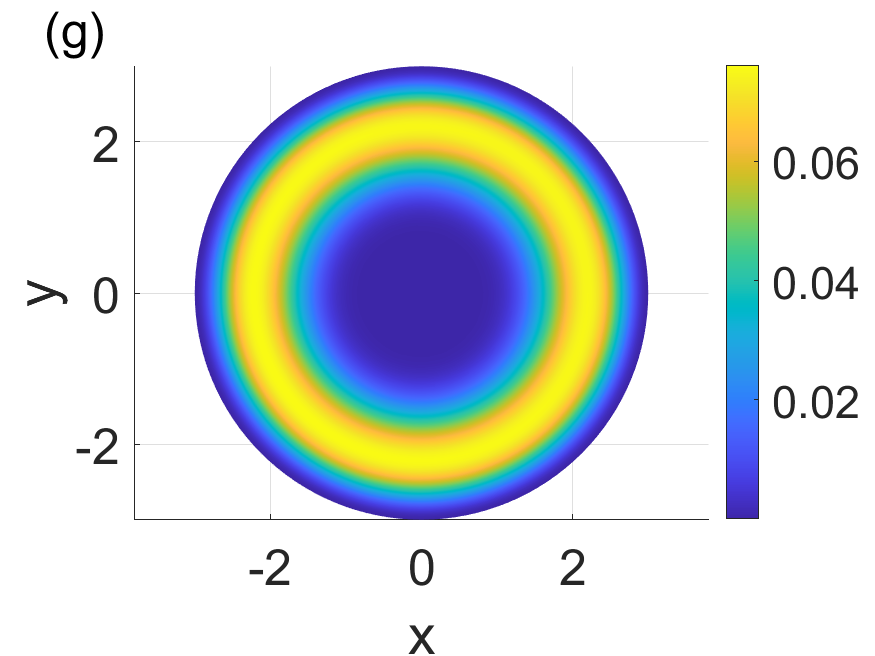}
\includegraphics[width=4.1cm,height=3.27cm,angle=-0]{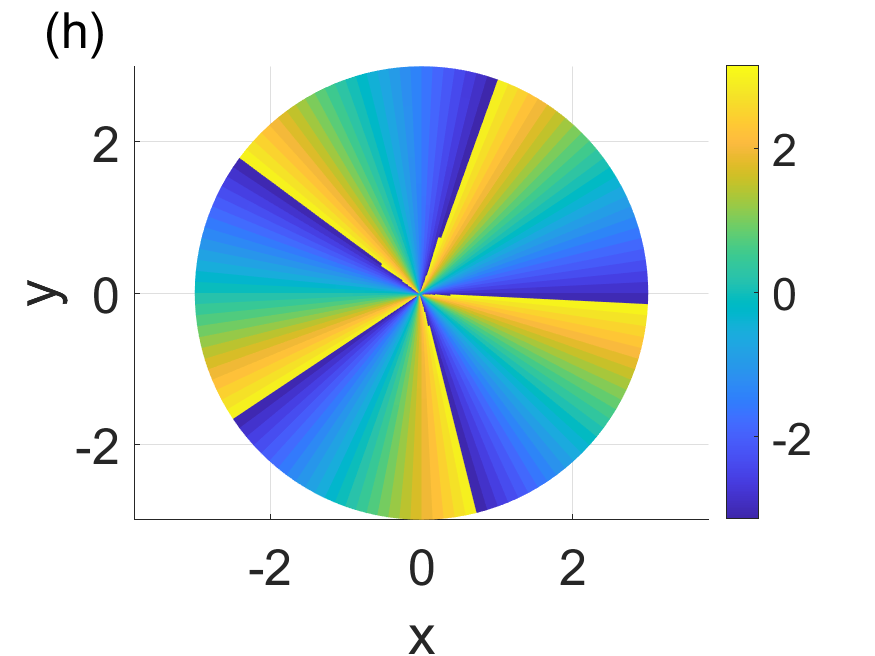}
\caption{(Colour online). (a), (c), (e), (g) Density and (b), (d), (f), (h) phase 
of the order parameter, in a hard-wall potential, with $R/R_0=3$, for the case of 
multiply-quantized vortex states, where the order parameter is $\propto 
e^{i m_0 \theta}$, with $g=2$. In (a) and (b) $\Omega/\Omega_0 = 0.75$, where 
$m_0=2$; in (c) and (d) $\Omega/\Omega_0 = 0.90$, where $m_0=3$; in (e) and (f) 
$\Omega/\Omega_0 = 1.05$, where $m_0=4$; in (g) and (h) $\Omega/\Omega_0 = 1.20$, 
where $m_0=5$. Here the axes are measured in units of $R_0$, and the density in 
units of $R_0^{-2}$.}
\end{figure}

\begin{figure}
\includegraphics[width=9cm,height=7cm,angle=-0]{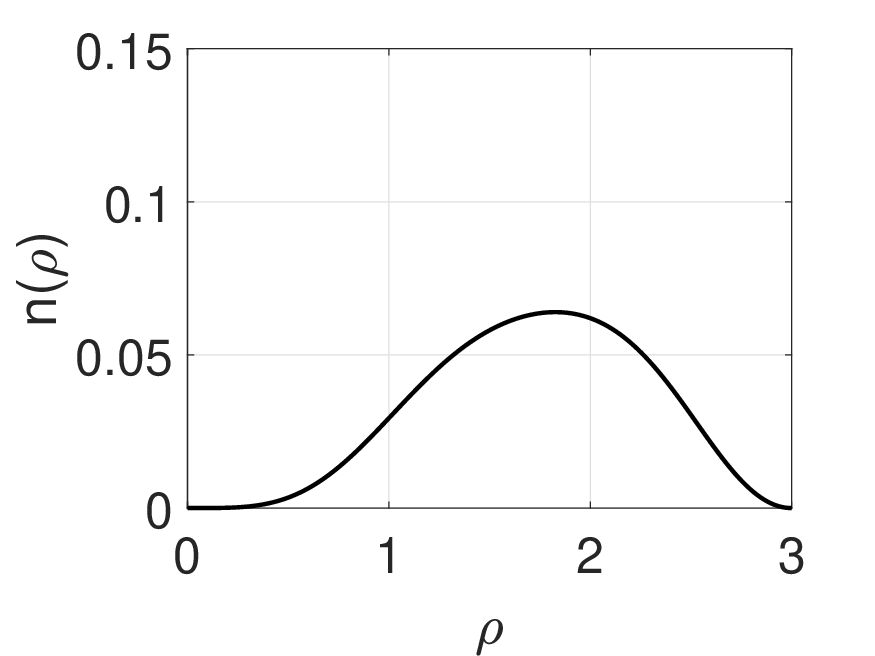}
\caption{The transverse profile of the density that corresponds to Fig.\,2 (a), i.e., 
for $g = 2$, $m_0 = 2$, and $\Omega/\Omega_0 = 0.75$. Here $R/R_0 = 3$, the $x$
axis is measured in units of $R_0$, and the $y$ axis in units of $R_0^{-2}$.}
\end{figure}

\begin{figure}
\includegraphics[width=4.1cm,height=3.27cm,angle=-0]{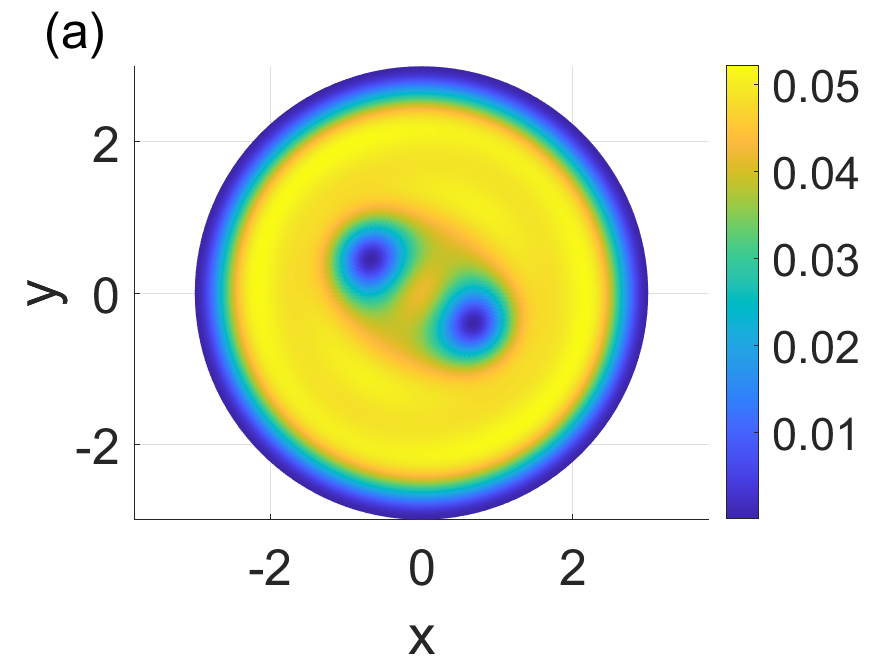}
\includegraphics[width=4.1cm,height=3.27cm,angle=-0]{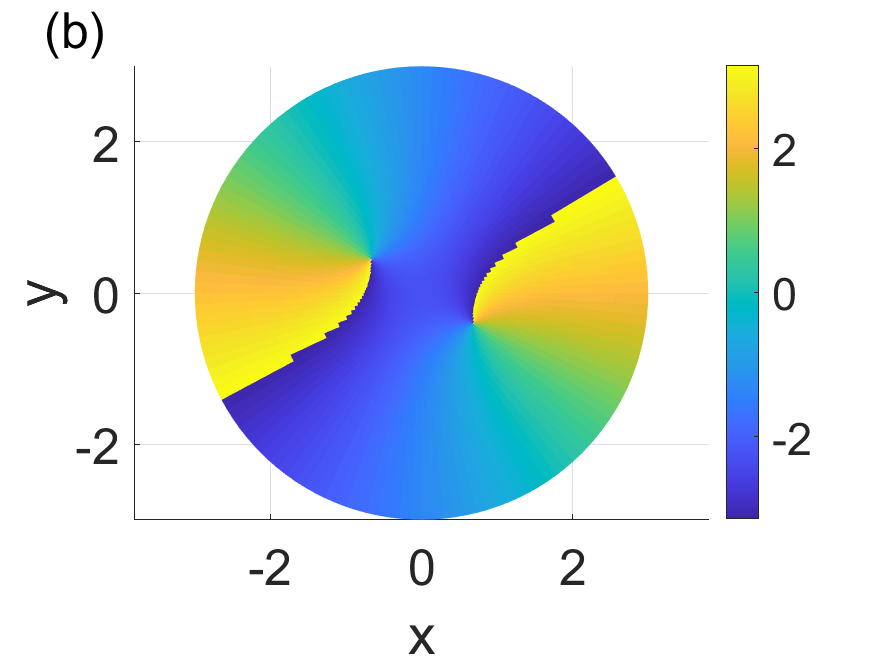}
\includegraphics[width=4.1cm,height=3.27cm,angle=-0]{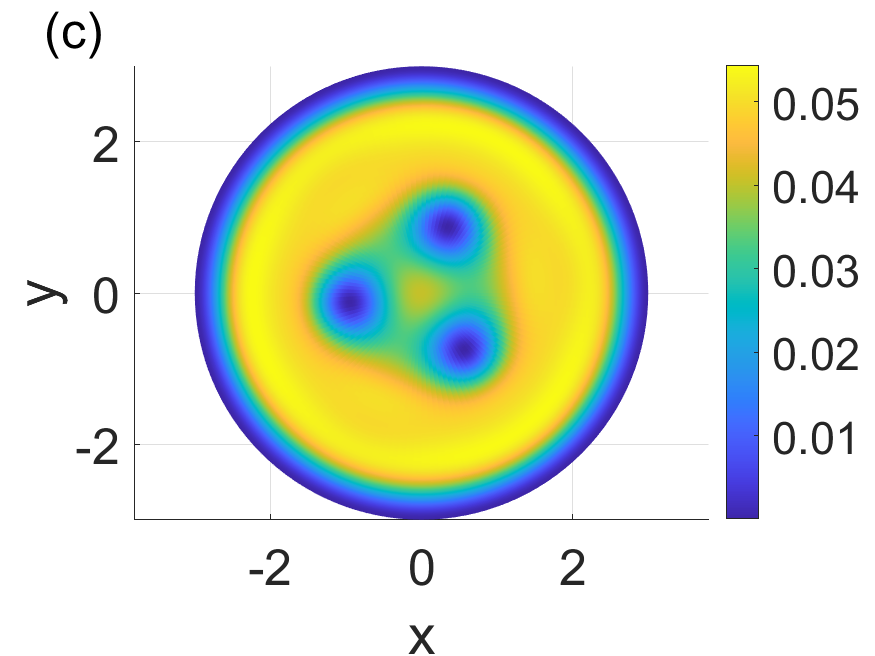}
\includegraphics[width=4.1cm,height=3.27cm,angle=-0]{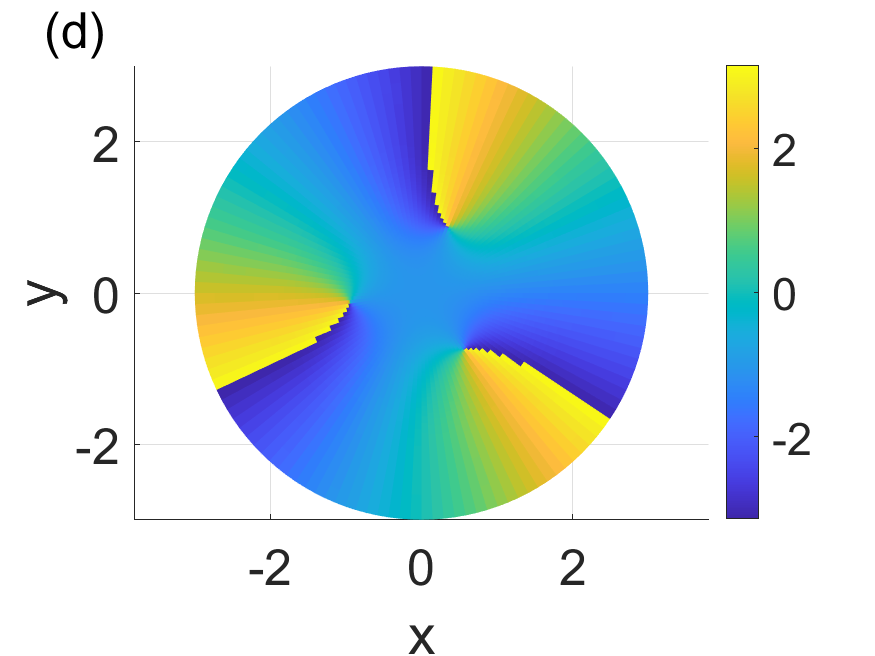}
\includegraphics[width=4.1cm,height=3.27cm,angle=-0]{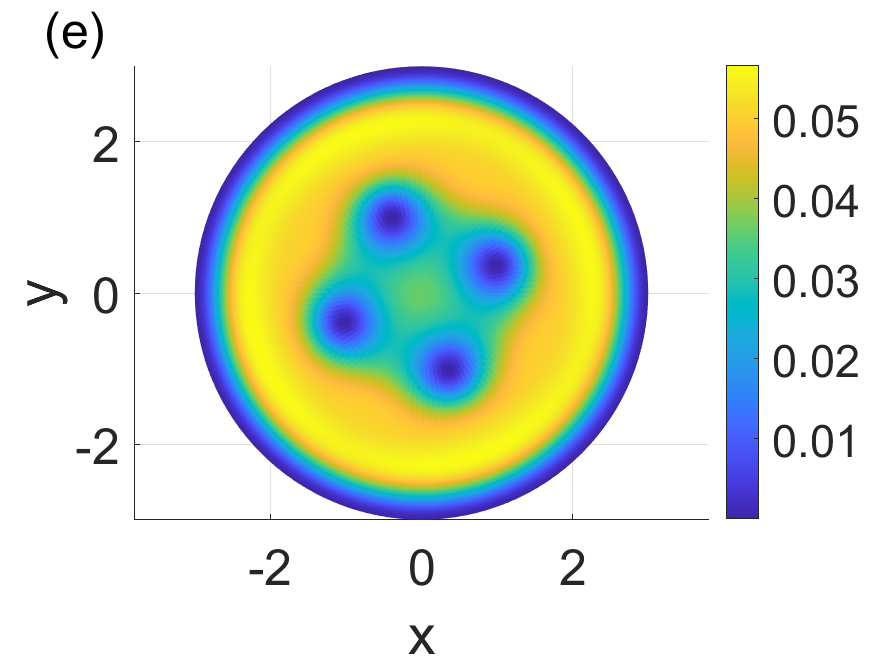}
\includegraphics[width=4.1cm,height=3.27cm,angle=-0]{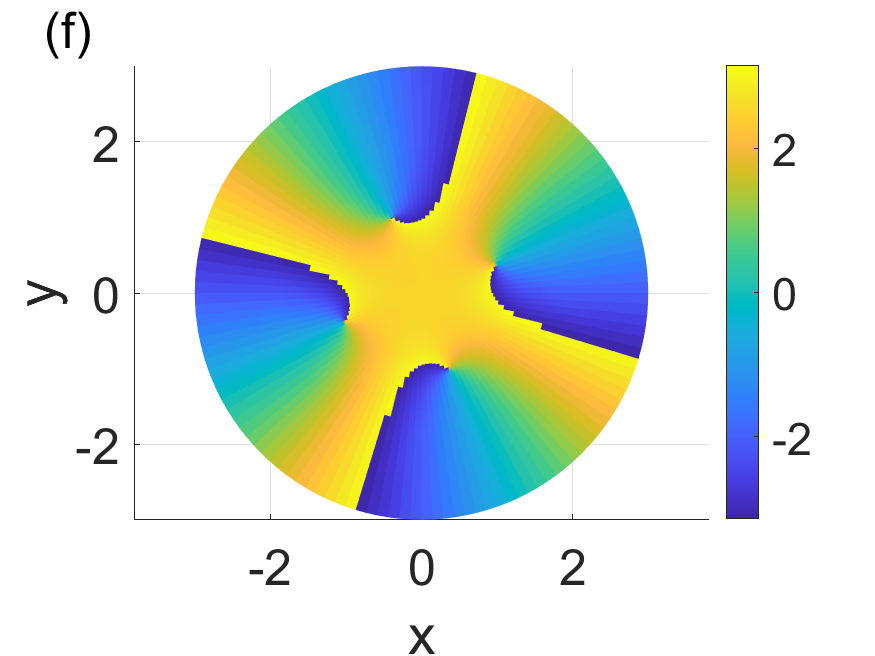}
\includegraphics[width=4.1cm,height=3.27cm,angle=-0]{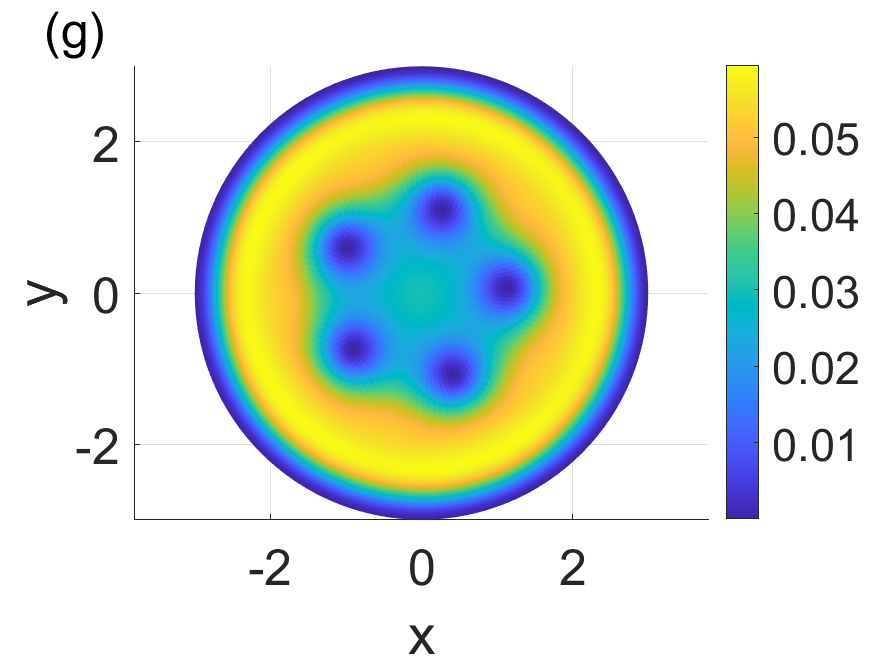}
\includegraphics[width=4.1cm,height=3.27cm,angle=-0]{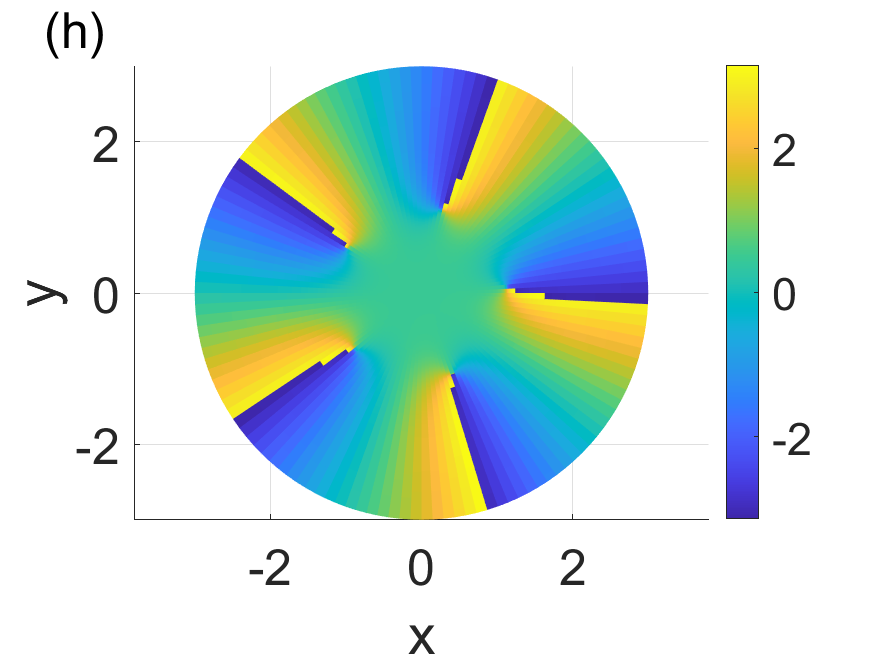}
\caption{(Colour online). (a), (c), (e), (g) Density and (b), (d), (f), (h) phase 
of the order parameter, in a hard-wall potential, with $R/R_0=3$, and $g=10$. In 
(a) and (b) $\Omega/\Omega_0 = 0.75$; in (c) and (d) $\Omega/\Omega_0 = 1.00$; in 
(e) and (f) $\Omega/\Omega_0 = 1.15$; in (g) and (h) $\Omega/\Omega_0 = 1.30$. Here 
the axes are measured in units of $R_0$, and the density in units of $R_0^{-2}$.}
\end{figure}

\section{Phase diagram}

\subsection{Zero, or weak interactions}

Let us start with the case of zero interactions, where one has the single-particle 
problem
\begin{eqnarray}
  - \frac {1} {2} \nabla^2 \psi(\rho, \theta) = \epsilon \psi(\rho, \theta),
\end{eqnarray}
with the boundary condition $\psi(\rho = R, \theta) = 0$. The above equation takes 
the form
\begin{eqnarray}
  \nabla^2 \psi(\rho, \theta) + k^2 \psi(\rho, \theta) = 0,
\label{eqqq}
\end{eqnarray}
with $k^2 = 2 \epsilon$. From Eq.\,(\ref{eqqq}) it follows that
\begin{eqnarray}
  \psi_{\rho \rho}(\rho, \theta) + \frac 1 {\rho} \psi_{\rho}(\rho, \theta)
  + \frac 1 {\rho^2} \psi_{\theta \theta}(\rho, \theta) 
  + k^2 \psi(\rho, \theta) = 0.
\nonumber \\
\end{eqnarray}
With the assumption that $\psi(\rho, \theta) = f(\rho) e^{i m \theta}$,
with $m$ being an integer that corresponds to the angular momentum,
\begin{eqnarray}
  {f}_{\rho \rho}(\rho) + \frac 1 {\rho} {f}_{\rho}(\rho)
  - \frac {m^2} {\rho^2} {f}(\rho) + k^2 {f}(\rho) = 0.
\end{eqnarray}
Therefore, the functions $f(\rho)$ are Bessel functions $J_{m}(k \rho)$, while 
the eigenfunctions (which are characterized by two quantum numbers, $n$ and $m$), 
are 
\begin{equation}
    \psi_{n,m}(\rho, \theta) = N_{n,m} J_m(k_{n,m} \rho) e^{i m \theta},
\end{equation}
where $N_{n,m}$ are the normalization constants. The boundary condition due to the 
assumption of a hard-wall potential $J_m(k_{n,m} R) = 0$ implies that $k_{n,m} R = 
\alpha_{n,m}$, (with $n = 0, 1, \dots )$, where $\alpha_{n,m}$ are the 
roots of $J_m$, and therefore $k_{n,m} = \alpha_{n,m}/R$. The single-particle 
energy spectrum is thus,
\begin{equation}
  \epsilon_{n,m} = \frac {\alpha_{n,m}^2} {2 R^2}. 
\end{equation}
In the above expressions $n$ is the quantum number that gives the radial nodes and $m$
is the angular momentum. Since the hard-wall potential goes more rapidly to infinity
than the harmonic potential, $\epsilon_{n,m}$ has a positive curvature as a function of
$m$ (for some given $n$). Including the interactions perturbatively, and setting $n=0$,
since all states with $n>0$ have higher energy, the total energy per particle in the 
rotating frame is \cite{e11},
\begin{eqnarray}
  \frac {E_m^{\rm rot}} N = \frac {\alpha_{0,m}^2} {2 R^2} - m \Omega 
  + 2 \pi g \int |\psi_{0,m}|^4 \, d^2 \rho,
  \label{erottot2}
\end{eqnarray}
with $g$ being either positive (i.e., an effective repulsive interaction), or negative
(i.e., an effective attractive interaction). In the limit of weak interactions that 
we consider, the single-particle energy is the dominant term. However, we argued above 
that the single-particle energy has a positive curvature as a function of $m$, therefore 
$E_m^{\rm rot}/N$ also has a positive curvature as a function of $m$. The important 
conclusion that follows is that for any fixed value of $\Omega$, there will be a 
unique value $m_0$ which minimizes $E_m^{\rm rot}/N$. 

Furthermore, as $\Omega$ increases, there are discontinuous transitions between 
states with winding number $m$ and $m+1$. The phase boundary that separates
these states on the $\Omega - g$ plane may be found by setting $E_{m+1}^{\rm rot} = 
E_{m}^{\rm rot}$. From Eq.\,(\ref{erottot2}) it follows that \cite{e11}
\begin{eqnarray}
  \Omega_{m, m+1} = \frac {\alpha_{0,m+1}^2 - \alpha_{0,m}^2} {2 R^2} +
\nonumber \\
  + 2 \pi g \int [|\psi_{0,m+1}|^4 - |\psi_{0,m}|^4] \, d^2 \rho.
  \label{neqq}
\end{eqnarray}
The last equation defines straight lines on the $\Omega - g$ plane, which are shown 
in Figs.\,1 and 8. 

To confirm the above result, we have performed extended numerical simulations. For
example, Fig.\,2 shows the density and the phase of the order parameter that we have
evaluated fully numerically (regarding the numerical techniques, see below) for the 
four points which are shown as circles in Fig.\,1. Indeed, Fig.\,2 shows 
multiply-quantized vortex states (which are rather close to the lowest Bessel 
eigenstates) with the expected value of circulation. Figure 3 shows the transverse 
profile of the density for one of the states of multiple quantization (left circle 
in Fig.\,1, with winding number equal to 2), which we also evaluate numerically. 

As we argued above, as the interaction strength $g$ increases, there is an instability 
of the states $\psi_{0,m}$ towards a mixed state, i.e., a linear superposition of the 
form \cite{e11}
\begin{equation}
  \Psi = c_0 \psi_{0,0} + c_m \psi_{0,m} + c_{2m} \psi_{0,2m}.
  \label{mixed}
\end{equation} 
The dashed lines in Fig.\,1 show the phase boundaries that were evaluated essentially 
analytically (i.e., diagonalizing matrices of dimension equal to 2) in Ref.\,\cite{e11}.
These boundaries separate the states of multiple quantization from the mixed states of 
the form of Eq.\,(\ref{mixed}). The points denoted by asterisks in Fig. 1 correspond 
to mixed states. Figure 4 shows the density and the phase of the order parameter that 
we evaluate numerically for these four points, confirming, again, the analytic result 
of Eq.\,(\ref{mixed}) \cite{e11}. We stress that comparing Figs.\,2 and 4 one sees 
immediately the expected instability of multiply-quantized vortex states towards mixed 
states.

Thus, the numerical simulations that we have performed not only confirmed the analytic 
results of Ref.\,\cite{e11}, but also allowed us to investigate the global phase diagram, 
which includes higher (positive) values of the coupling $g$, as well as negative values 
of $g$. These results are presented below. 

Regarding our numerical approach, we have adopted two different strategies, both of which 
are based on the method of imaginary-time propagation \cite{imt}. More specifically, we 
minimize the following energy functional in the rotating frame, for some fixed atom number 
and some fixed value of $\Omega$,
\begin{eqnarray}
  {\cal E}(\Psi, \Psi^*) = \frac 1 2 \int |\nabla \Psi|^2 \, d^2 \rho 
  + 2 \pi g \int |\Psi|^4 \, d^2 \rho 
  \nonumber \\
  - \mu \int |\Psi|^2 \, d^2 \rho - \Omega 
  \int \Psi^* {\hat L} \Psi \, d^2 \rho, 
  \label{functional}
\end{eqnarray}
where ${\hat L}$ is the angular momentum operator. The confining potential is taken into 
account via the boundary condition $\Psi(\rho=R,\theta) = 0$. 

Within the first approach, we expand the order parameter in the basis of the Bessel 
eigenfunctions. This approach is more suitable for small interaction strengths and this 
is where we use it in the present study. Within the second approach we start with a 
reasonable initial state and propagate it in imaginary time, without making any further 
assumptions about the form of the order parameter. We use this approach for the case of 
moderate to strong couplings. 

The choice of initial conditions in both approaches requires special attention, 
since one has to avoid getting trapped in some local minimum, instead of finding the 
global minimum of the energy functional. To minimize the possibility of convergence 
to a metastable state, we have used various initial conditions and have also used the 
available analytic results as a guide. Regarding the convergence, the criterion 
that we have used is based on the energy, which we monitor as a function of the 
imaginary time. The number of Bessel states that we use depends on the values 
of $g$ and $\Omega$, however we have confirmed that the results are unchanged when 
the basis size increases. Finally, as mentioned above, the hard-wall boundary 
is implemented numerically by demanding that the order parameter vanishes at the boundary.

\subsection{Thomas-Fermi limit and more general phase diagram for repulsive interactions}

Let us now turn to the more general problem, where the interaction is no longer weak.
The numerical approach that we described above allows us to investigate this problem, 
as well. The derived results are shown in the phase diagram of Fig.\,5 (a log-log
plot), in the density and phase plots of Fig.\,6 and in the density profile of Fig.\,7.
In the left corner on the bottom of the phase diagram of Fig.\,5 we include the points
of Fig.\,1. For higher values of $g$ and $\Omega$, the phase diagram includes three
phases, namely a vortex lattice (squares), see, also, Figs.\,6 (a) to (h), a vortex 
lattice with a hole (stars), see, also, Figs.\,6 (i) to (l), and finally a giant 
vortex (diamonds), see, also, Figs.\,6 (m) and (n). 
 
To get some insight, let us turn to the Thomas-Fermi limit,
$g \gg 1$. Here we follow the variational approach by Fischer and Baym \cite{e4} 
to derive some analytic results, in an effort to compare them with our numerical data. 
These authors considered the limit of sufficiently rapid rotation, in which a vortex 
lattice has formed. The main idea is that, in the presence of a vortex lattice there 
is a smoothed, slowly varying, density distribution $n_{\rm WS}({\rho})$ and, on top 
of that, a rapidly varying density distribution due to the presence of the vortices. 

Among other interesting results, Fischer and Baym found that $n_{\rm WS}({\rho})$ develops
a hole at the center of the cloud, due to the effect of the centrifugal potential. To see 
that, let us start with the following expression for the (two-dimensional) density 
$n_{\rm WS}(\rho)$ \cite{e4},
\begin{equation}
   n_{\rm WS}(\rho) = \frac 1 {g b} \frac {N} {4 \pi} 
   \left(\mu - a \Omega + \frac 1 2 \Omega^2 \rho^2 \right).
\label{dens1} 
\end{equation}
Here $a$ and $b$ are variational parameters, and $\mu$ is the chemical potential. Since 
$n_{\rm WS}(\rho = 0) \propto \mu - a \Omega$, the condition for the formation of a hole 
is that $\mu - a \Omega = 0$. Integrating $n_{\rm WS}(\rho)$ over the surface of the disk, 
from the conservation of the atom number, it follows that
\begin{equation}
   g = \frac 1 {16 b} {(\Omega R^2)^2}.
\label{gphb}
\end{equation}
Furthermore, $b$ varies between 1 and 4/3. In the Thomas-Fermi limit, where the healing 
length is small compared with the size of the lattice cell that corresponds to each 
vortex, $b=1$. In the more general phase diagram of Fig.\,5 that we have derived 
numerically, the result of Eq.\,(\ref{gphb}) is shown as the left dashed straight 
line ($g$ scales as $\Omega^2)$, for $R/R_0=3$ and $b=1$. As we see, this limiting 
analytic result is reasonably close to the phase boundary between a vortex lattice 
(squares) and a vortex lattice with a hole (stars).

For even higher values of $\Omega$, the vortex lattice disappears completely and a giant 
vortex forms \cite{e4}, as a result of the even stronger centrifugal potential. Denoting 
the radius of the hole as $R_G$ and setting $R_G = \alpha R$, with $\alpha < 1$, the 
smoothed density distribution is given by
\begin{equation}
   n_{\rm WS}(\rho) = \frac {N m_0^2} {8 \pi g} 
   \left( \frac 1 {R_G^2} - \frac 1 {\rho^2} \right). 
\label{dens2}
\end{equation}

Clearly the density in the above equation has to vanish for $\rho = R_G$, and $\rho = R$ 
(see, e.g., Fig. 7). The first condition is satisfied, the second is not. Still, the 
difference in the energy between a solution that satisfies the second condition and 
the one in Eq.\,(\ref{dens2}) differs by the kinetic energy associated with the bending 
of the order parameter, which is negligible in the assumed Thomas-Fermi limit. In 
Eq.\,(\ref{dens2}) $m_0$ is the winding number of the giant vortex, which in the limit 
of rapid rotation is $\approx \Omega R^2$, as Feynman's formula implies. Again, 
integrating over the surface of the disk we get that
\begin{equation}
  g = \frac 1 4 (\Omega R^2)^2  
  \left[ \frac 1 2 \left( \frac 1 {\alpha^2} - 1 \right) + \ln \alpha \right].
\label{ggg}
\end{equation}
In the limit of rapid rotation, $\alpha \to 1$ and therefore Eq.\,(\ref{ggg}) implies that
\begin{equation}
  g \approx \frac 1 4 (\Omega R^2)^2 (1 - \alpha)^2.
\label{gggg}
\end{equation}
To get an estimate about the location of the phase boundary predicted by 
Eq.\,(\ref{gggg}), we set $\alpha = 2.17/3.00$, from the numerical data of Fig.\,7. 
Using this value, we plot in Fig.\,5 Eq.\,(\ref{gggg}) (right dashed line). Again, 
the phase boundary is in reasonable agreement with the numerical results (stars vs 
diamonds). On the other hand, we stress that this is just an approximate result, since
a full calculation would require that the dependence of $\alpha$ on $g$ and $\Omega$
is also evaluated and included.

Some representative states obtained numerically from the phase diagram of Fig.\,5 are 
shown in Figs.\,6 and 7. In Fig.\,6 we have considered a value of $g=200$ 
and various values of $\Omega$. There, a vortex lattice is seen clearly, which eventually 
turns into a vortex lattice with a hole and finally into a giant vortex. Figure 7 shows 
the transverse density profile of a giant vortex. As seen in this plot, the strong 
centrifugal potential pushes the cloud towards the edge of the hard-wall boundary. 
Although this phase resembles the one of multiple quantization (see Fig.\,3), it has 
a rather distinct density distribution. The basic difference is that the state of 
Fig.\,7 is dominated by the interaction energy, whereas that of Fig.\,3 is dominated 
by the kinetic energy. In the giant-vortex state of Fig.\,7 the density is pushed 
strongly toward the hard-wall boundary, whereas the multiply-quantized vortex state 
of Fig.\,3 has a broader radial profile and the density extends further inside the disk.

\begin{figure}
\includegraphics[width=9cm,height=7cm,angle=-0]{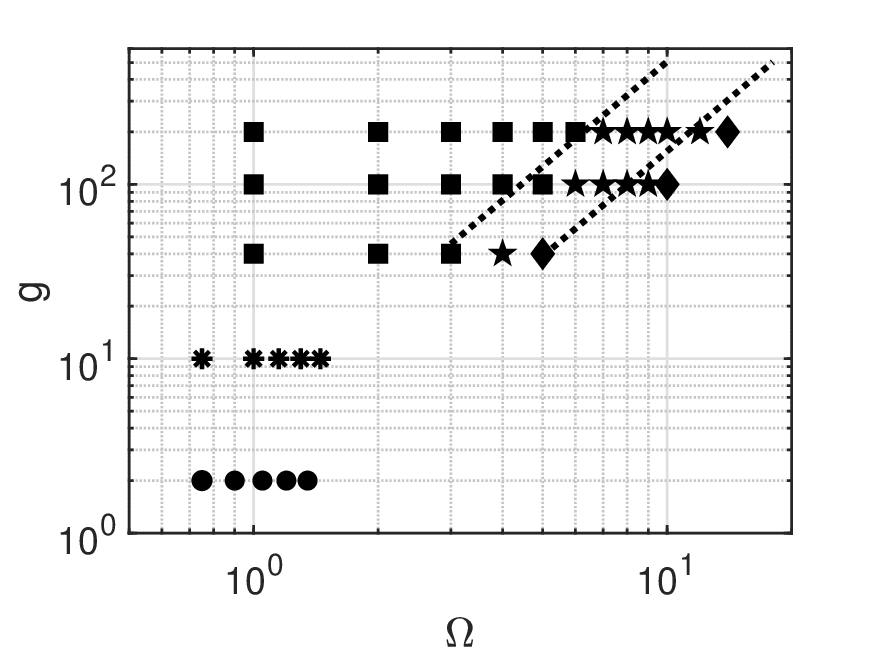}
\caption{More general phase diagram (log-log plot), where on the $x$ axis is the 
rotational frequency of the trap $\Omega$ (in units of $\Omega_0$) and on the $y$ 
axis is the (dimensionless) coupling $g$. Here we have a hard-wall trapping potential, 
with $R/R_0 = 3$. The circles ``$\bullet$" denote the multiply-quantized vortex states, 
which are very close to the Bessel eigenstates of the non-interacting problem. The 
asterisks ``${\bf \ast}$" denote the mixed states, which are linear superpositions of Bessel 
eigenstates. The squares ``$\blacksquare$" denote the phase of a vortex lattice. The 
stars ``$\bigstar$" denote the phase of a vortex lattice with a hole. Finally, the 
diamonds ``$\blacklozenge$" denote the giant-vortex phase, which resembles the  
multiply-quantized vortex states (circles). The left dashed line is Eq.\,(\ref{gphb}), 
while the right one is Eq.\,(\ref{gggg}).}
\end{figure}

\begin{figure}
\includegraphics[width=4.1cm,height=3.27cm,angle=-0]{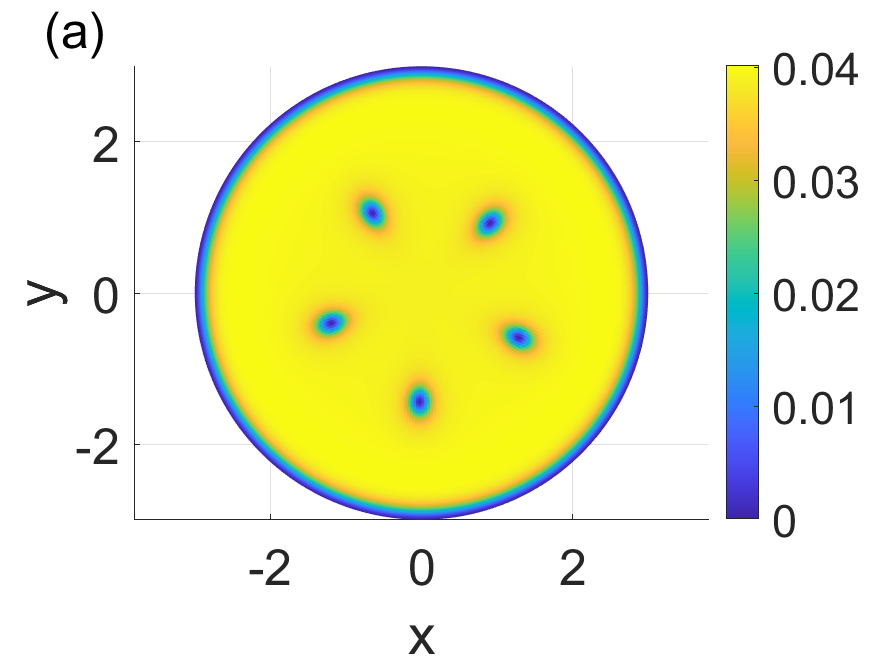}
\includegraphics[width=4.1cm,height=3.27cm,angle=-0]{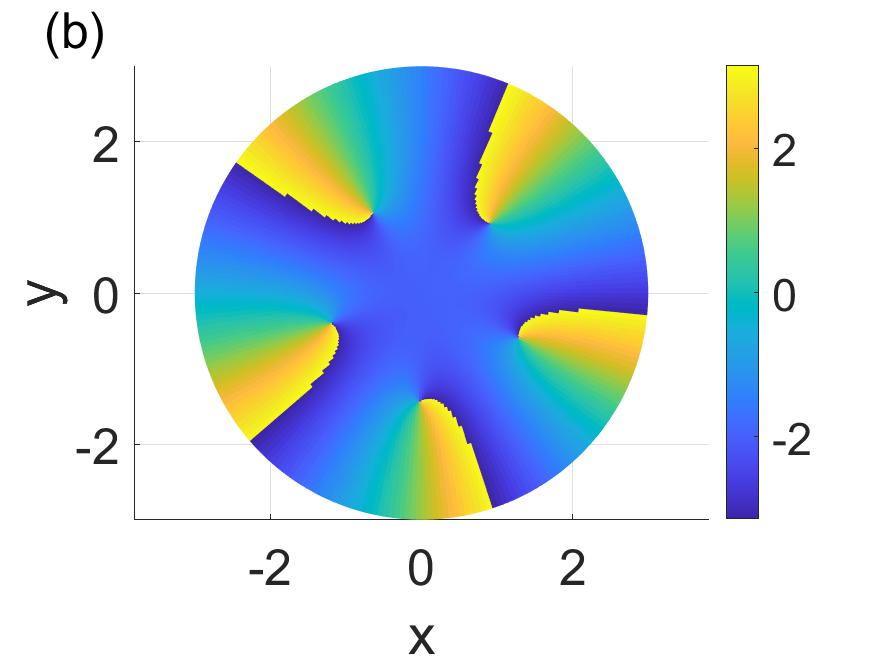}
\includegraphics[width=4.1cm,height=3.27cm,angle=-0]{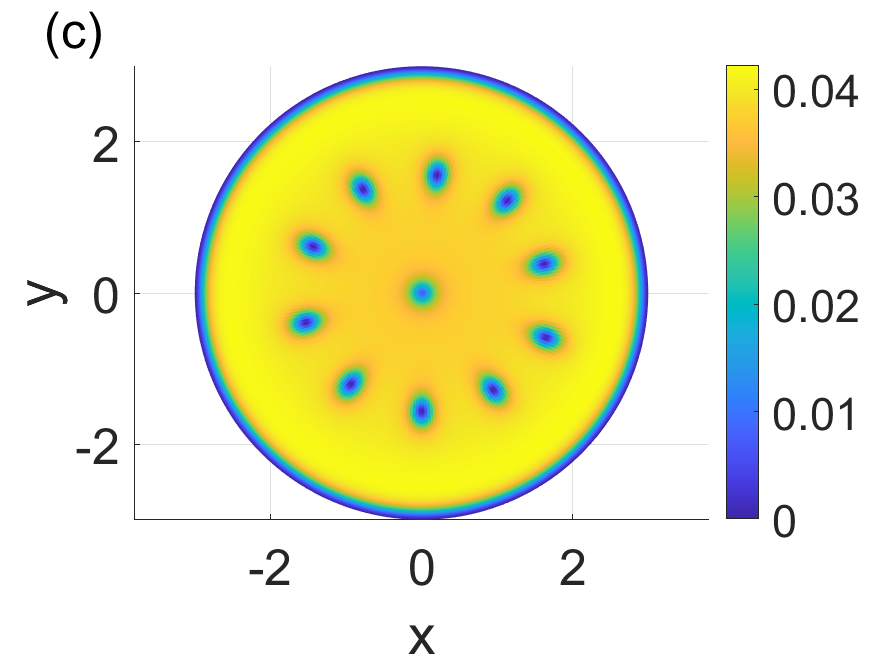}
\includegraphics[width=4.1cm,height=3.27cm,angle=-0]{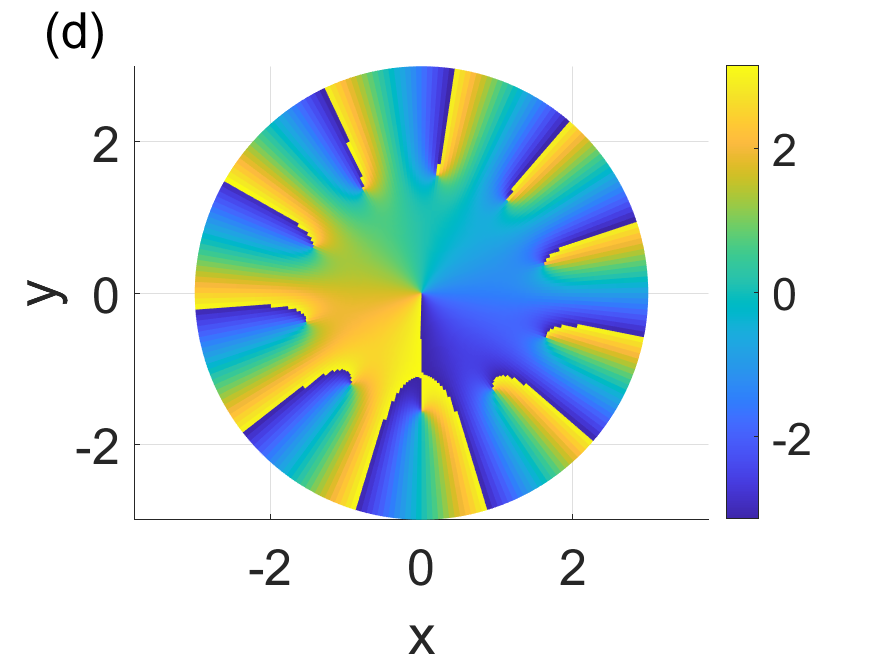}
\includegraphics[width=4.1cm,height=3.27cm,angle=-0]{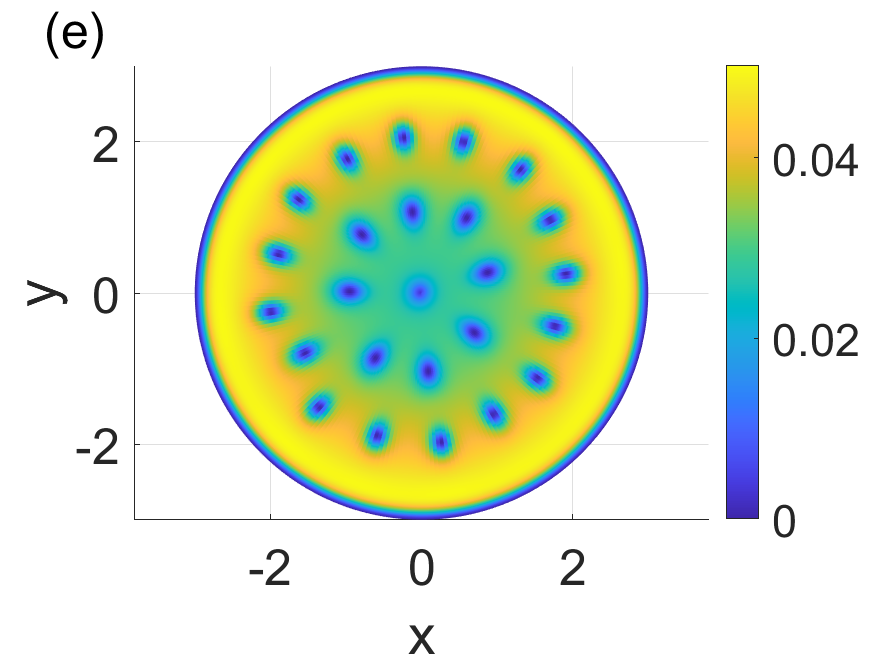}
\includegraphics[width=4.1cm,height=3.27cm,angle=-0]{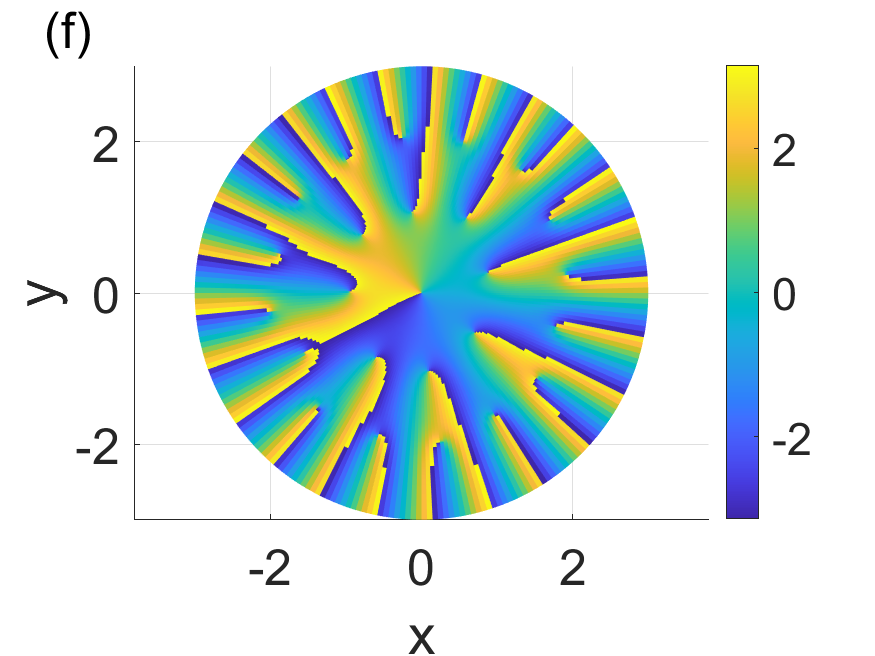}
\includegraphics[width=4.1cm,height=3.27cm,angle=-0]{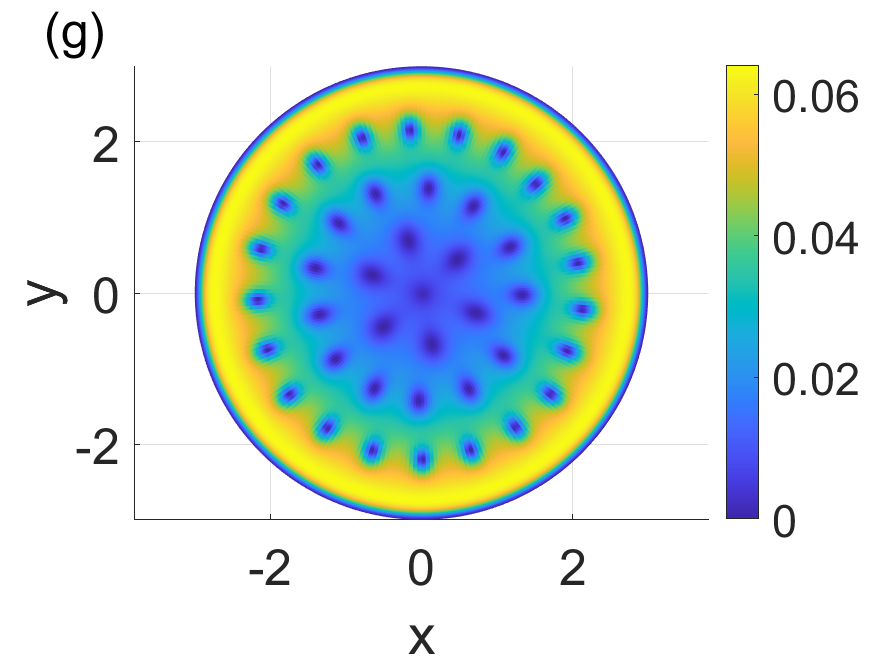}
\includegraphics[width=4.1cm,height=3.27cm,angle=-0]{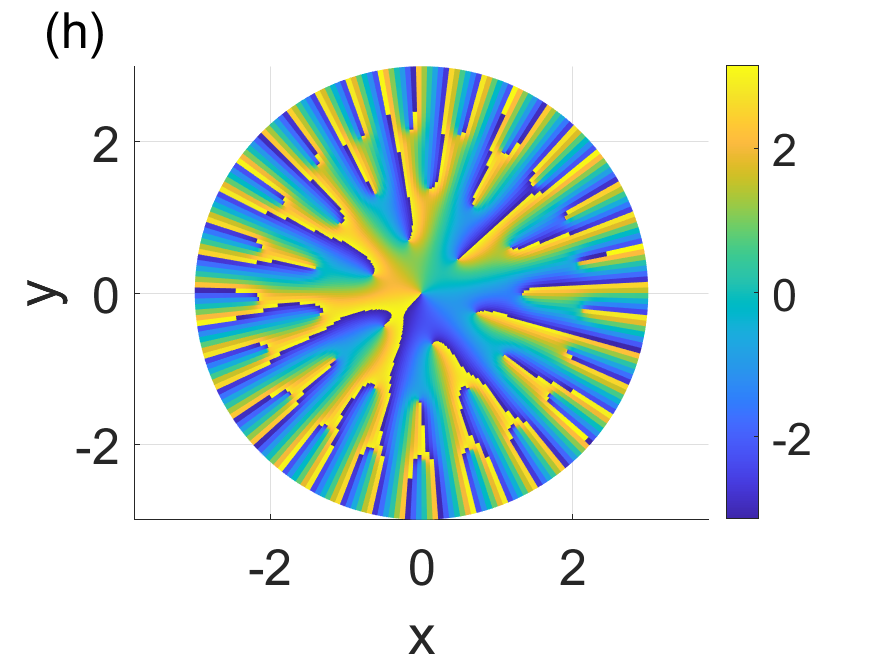}
\includegraphics[width=4.1cm,height=3.27cm,angle=-0]{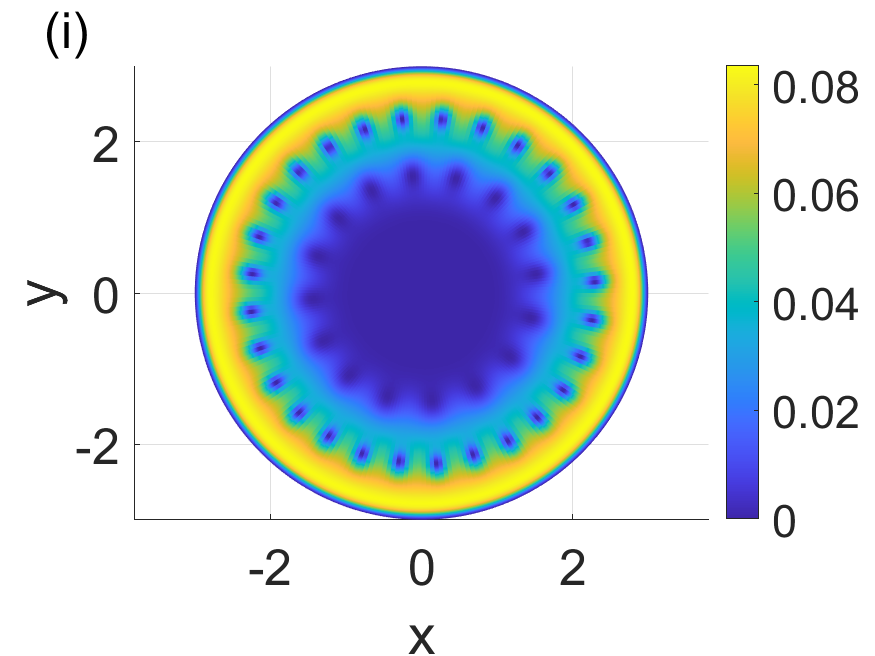}
\includegraphics[width=4.1cm,height=3.27cm,angle=-0]{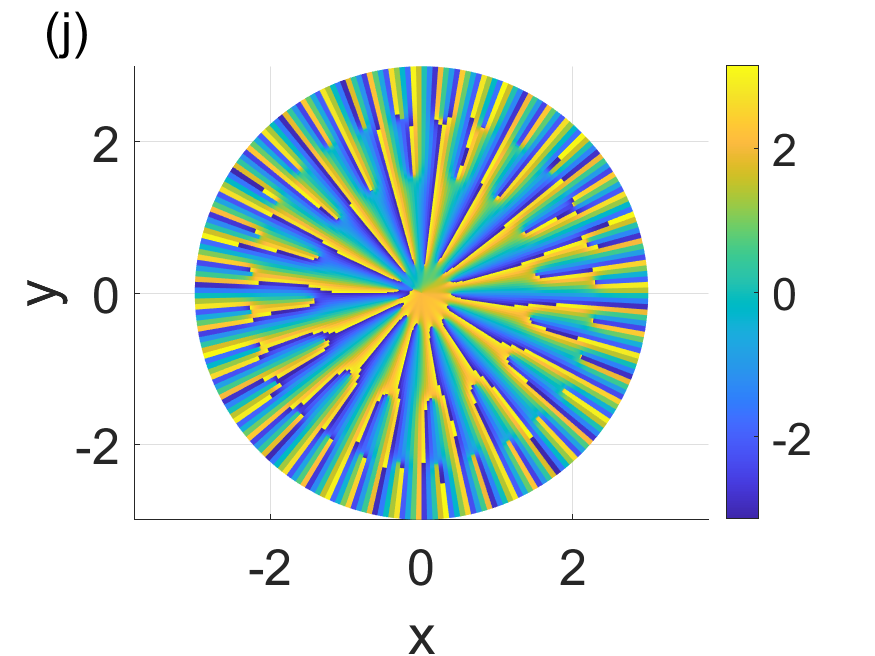}
\includegraphics[width=4.1cm,height=3.27cm,angle=-0]{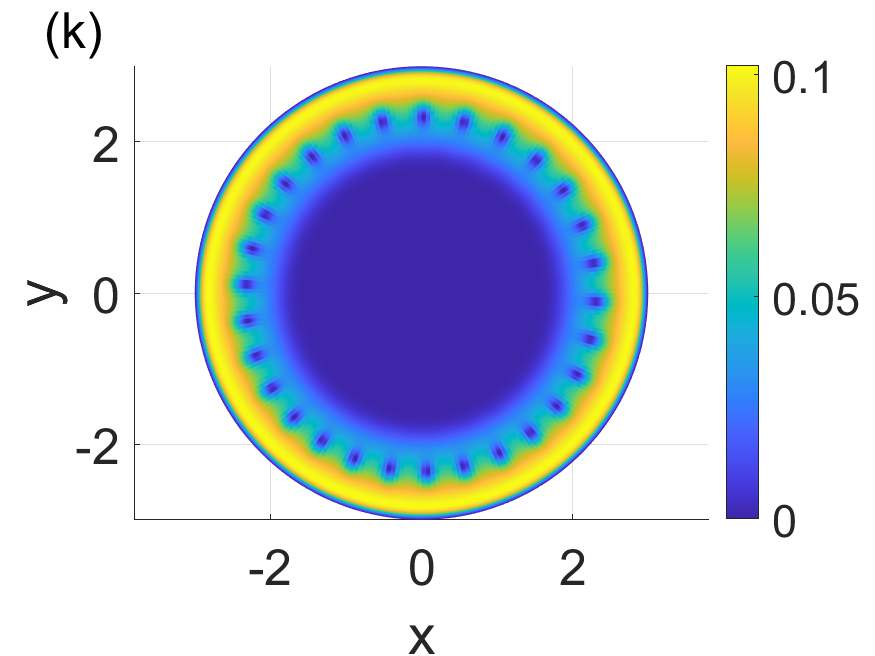}
\includegraphics[width=4.1cm,height=3.27cm,angle=-0]{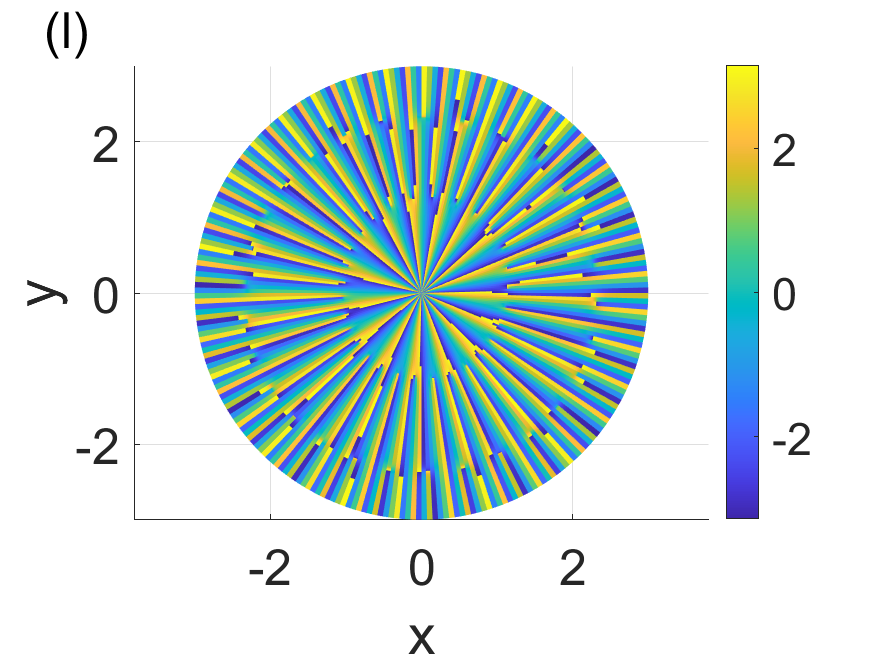}
\includegraphics[width=4.1cm,height=3.27cm,angle=-0]{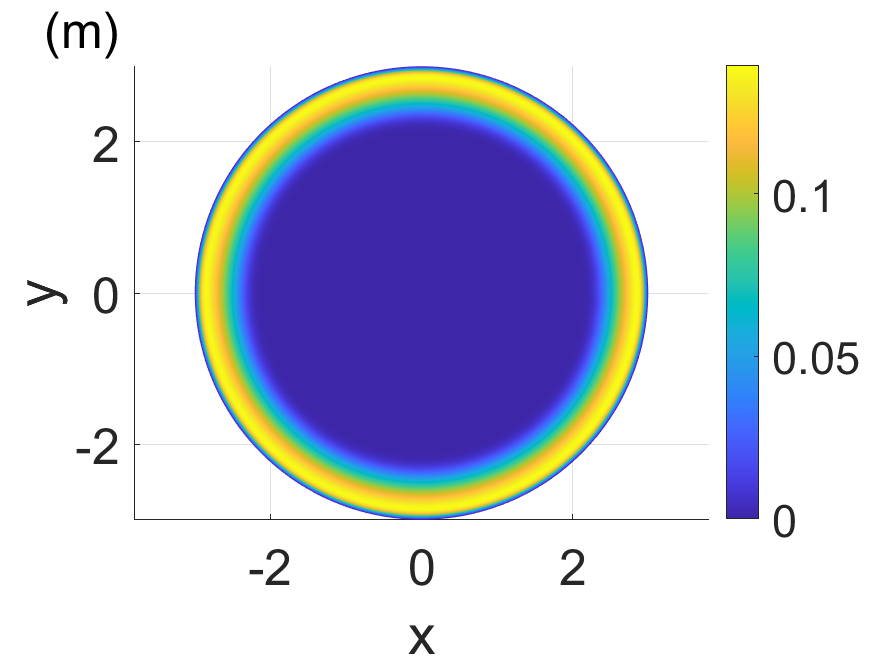}
\includegraphics[width=4.1cm,height=3.27cm,angle=-0]{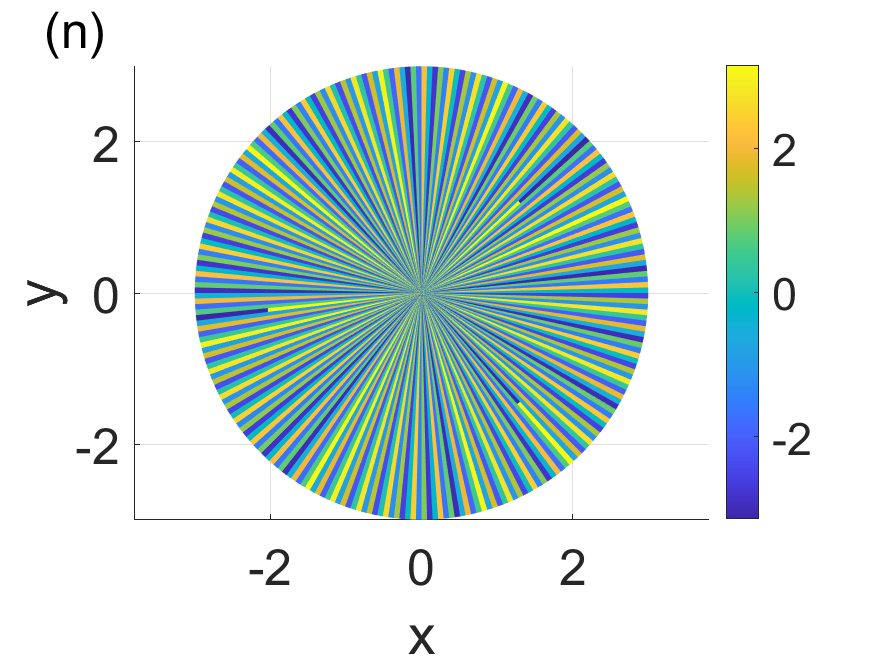}
\caption{(Colour online). (a), (c), (e), (g), (i), (k), (m) Density and 
(b), (d), (f), (h), (j), (l), (n) phase of the order parameter, in a hard-wall 
potential, with $R/R_0=3$, and $g=200$. In (a) and (b) $\Omega/\Omega_0 = 1$; 
in (c) and (d) $\Omega/\Omega_0 = 2$; in (e) and (f) $\Omega/\Omega_0 = 4$; 
in (g) and (h) $\Omega/\Omega_0 = 6$; in (i) and (j) $\Omega/\Omega_0 = 8$; 
in (k) and (l) $\Omega/\Omega_0 = 10$; in (m) and (n) $\Omega/\Omega_0 = 14$. 
Here the axes are measured in units of $R_0$, and the density in units of 
$R_0^{-2}$.}
\end{figure}

\begin{figure}
\includegraphics[width=9cm,height=7cm,angle=-0]{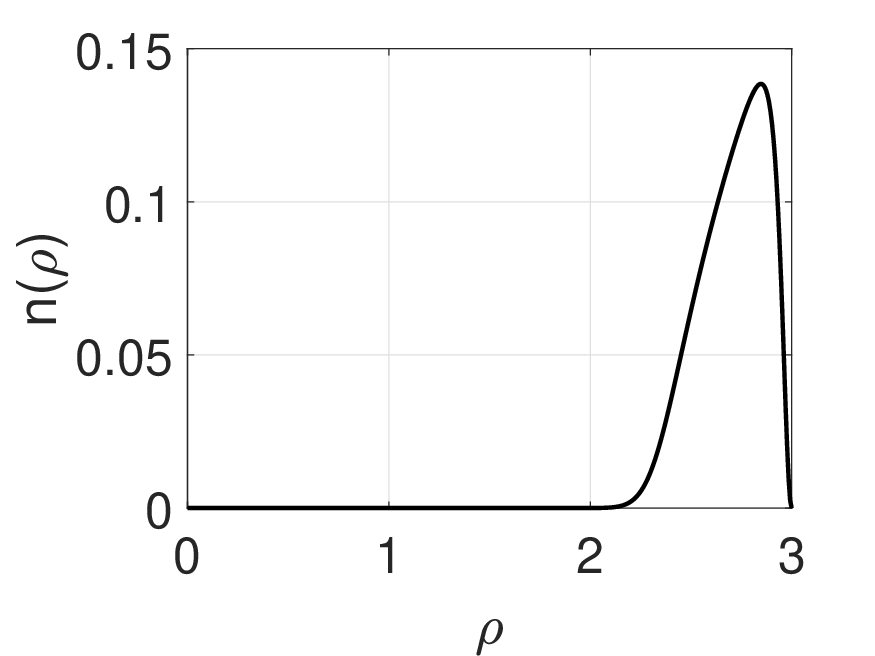}
\caption{The transverse profile of the density that corresponds to Fig.\,6 (m), 
for $g = 200$, 
and $\Omega/\Omega_0 = 14$. Here $R/R_0 = 3$, the $x$
axis is measured in units of $R_0$, and the $y$ axis in units of $R_0^{-2}$.}
\end{figure}

\section{Phase diagram -- Effective attractive interactions}

While in free space a Bose-Einstein condensate with an effective attractive interaction
is unstable against collapse, in the present problem the system may be stabilized 
due to the presence of the hard-wall potential. Clearly the Thomas-Fermi limit is not 
accessible in this case, since if the kinetic energy is negligible and the interaction 
energy is dominant, there is no mechanism which will prevent the collapse of the system. 
Furthermore, giving angular momentum to the system, enhances its stability, as shown in 
the results that follow below, see Fig. 8.

Below we thus investigate the problem for an effective attractive interaction. The 
phase of multiple quantization that we discussed in Sec.\,III A is also present for 
sufficiently weak (attractive) interactions. The reason is that the dominant term in 
the energy is still the single-particle energy and as a result $E_m^{\rm rot}$ still 
has a positive curvature as a function of $m$ (see Eq.\,(\ref{erottot2})). This phase 
is shown in Fig.\,8 (circles). Also, the density and the phase of the order parameter 
for two specific such states (with a winding number equal to 1 and 4) are shown in 
Figs.\,9 (a) and (b) and Figs.\,10 (a) and (b), respectively.

As $g$ becomes more negative an instability sets in, where the density localizes 
and breaks axial symmetry. We refer to this state as center-of-mass-like because 
the density maximum is displaced from the symmetry axis. Such states are shown as 
diamonds in Figs.\,8. The density and the phase of the order parameter of two such 
examples are shown in Figs.\,9 (c) and (d), and 10 (c) and (d). 

Although the multiply-quantized state also possesses the mixed-state instability 
described by Eq.\,(\ref{mixed}) for $g < 0$, this instability does not give rise 
to a distinct ground-state phase in the hard-wall geometry. The enhancement of the 
instability towards the center-of-mass-like state in the hard-wall potential may 
be understood qualitatively from the effective potential in the rotating frame, 
$V_{\rm eff}(\rho) = V(\rho) - \Omega^2 \rho^2/2$. Inside the hard wall the 
trapping potential is constant, so $V_{\rm eff}(\rho)$ decreases monotonically 
towards the boundary. As a result, when the density localizes, it shifts outwards, 
which is energetically favourable. This differs from the quadratic-plus-quartic 
confinement \cite{KJB}, where the effective potential possesses a smooth minimum 
at a finite radial position. In that geometry, the mixed-state instability gives 
rise to a distinct phase, as shown in Ref.\,\cite{KJB}.

An analogous instability towards a localized state that breaks the axial symmetry of 
the Hamiltonian has been seen in a quasi-one-dimensional toroidal potential with an 
effective attractive interaction, see, e.g., Refs.\,\cite{Ueda, Kav}. Following these 
references, it turns out that we have formation of a localized blob (for zero angular 
momentum) if
\begin{equation}
  \frac {4 N a R_t} S < - \frac 1 2.
\end{equation}
Here $R_t$ is the mean radius of the torus and $S$ is its cross section. Assuming 
that the width of the torus is $R_t/3$ (see, e.g., Fig.\,3), the critical value of 
$g$ for the formation of a localized density distribution is $-\pi/24$. 

As $|g|$ increases further, the width of the blob decreases, and the peak density 
increases. While in one dimension there is no instability, in the present 
two-dimensional problem, above some critical value of $|g|$, the system is no 
longer stable and undergoes a collapse -- this phase is shown as the stars 
in Fig.\,8. One may perform an elementary variational calculation using the 
(normalized) lowest Bessel function $J_0(\alpha_{0,0}\rho/r_0)$ with some trial 
radius $r_0$. The choice of this trial function is justified, since we are 
interested in the instability of the zero angular-momentum state, i.e., $m = 0$, 
while any other Bessel state with $n \neq 0$ has a higher energy. The energy per 
particle is, in this state,
\begin{equation}
  \frac E N \approx \frac {2.89} {r_0^2} + \frac {4.20 g} {r_0^2}, 
  \label{crgg}
\end{equation} 
where the first term on the right is the kinetic energy and the second is
the interaction energy. The kinetic energy always scales as $1/r_0^2$. The 
fact that the interaction energy also scales as $1/r_0^2$ is due to the fact 
that we are in two spatial dimensions, see, e.g., Ref.\,\cite{Dal}. Equation 
(\ref{crgg}) implies that for $g > -0.689$ the energy is minimized by the 
largest allowed value, $r_0 = R$. For $g < -0.689$, however, the energy 
becomes unbounded from below as $r_0 \to 0$, signalling collapse. Therefore, 
the variational estimate for the onset of collapse in the static problem is 
$g_c = - 0.689$.

In the phase diagram of Fig.\,8 we observe that the critical value of $|g|$ 
increases with $\Omega$. As we see from this diagram, for some given $\Omega$, 
as $g$ becomes more negative first there is a phase transition to a localized 
state, before -- for even more negative values of $g$ -- the localized blob 
collapses. We should mention at this point that numerically the collapse shows 
up rather clearly. Looking at the amplitudes of the Bessel states which constitute 
the order parameter, in the stable phase these decay as the quantum numbers $n$ 
and $m$ increase. On the other hand, in the unstable phase, they increase. 

To conclude, the phase diagram of Fig.\,8 is, to some extent, similar to that 
found previously for a condensate in a quadratic-plus-quartic potential \cite{KJB}. 
On the one hand, the phase of multiple quantization, the center-of-mass-like phase
and the unstable phase are common in both problems. On the other hand, the mixed 
phase is absent from the phase diagram in the hard-wall potential for an effective
attractive interaction.

\begin{figure}
\includegraphics[width=9cm,height=7cm,angle=-0]{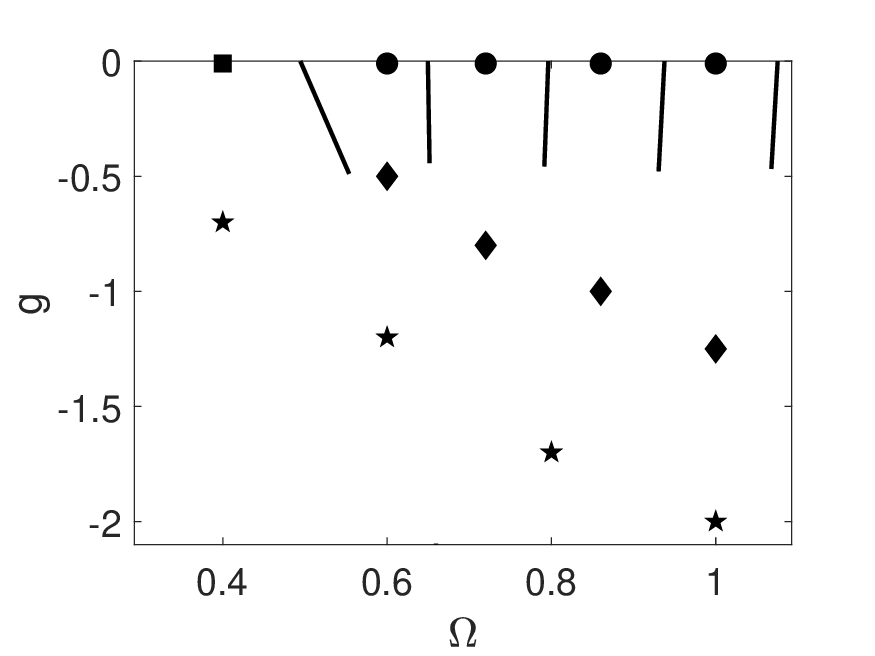}
\caption{Phase diagram, for effective attractive interactions, where on the $x$ axis 
is the rotational frequency of the trap $\Omega$ (in units of $\Omega_0$) and on 
the $y$ axis is the (dimensionless) coupling $g$. Here we have a hard-wall trapping 
potential, $R/R_0 = 3$. The solid lines, which are given by Eq.\,(\ref{neqq}), give 
the discontinuous transitions. In the region of multiple quantization, each time one 
crosses a solid line, the value of circulation increases by one unit, starting from 
zero. Here the circles ``$\bullet$" denote the states of multiple quantization, the 
square ``$\blacksquare$" denotes the static cloud, the diamonds ``$\blacklozenge$" 
denote the states with center-of-mass-like excitation, and the stars ``$\bigstar$" 
denote the unstable phase.}
\end{figure}

\begin{figure}
\includegraphics[width=4.1cm,height=3.27cm,angle=-0]{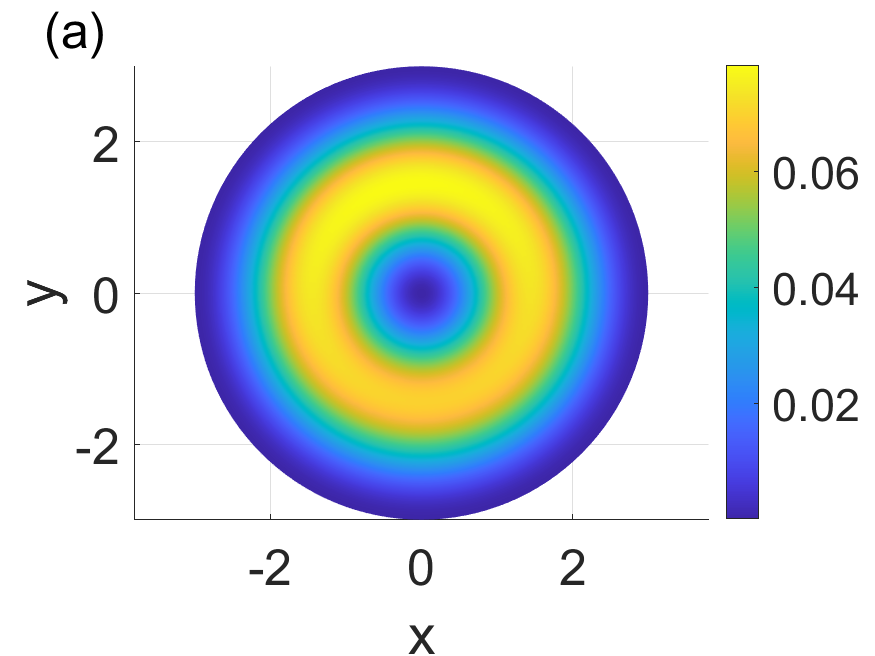}
\includegraphics[width=4.1cm,height=3.27cm,angle=-0]{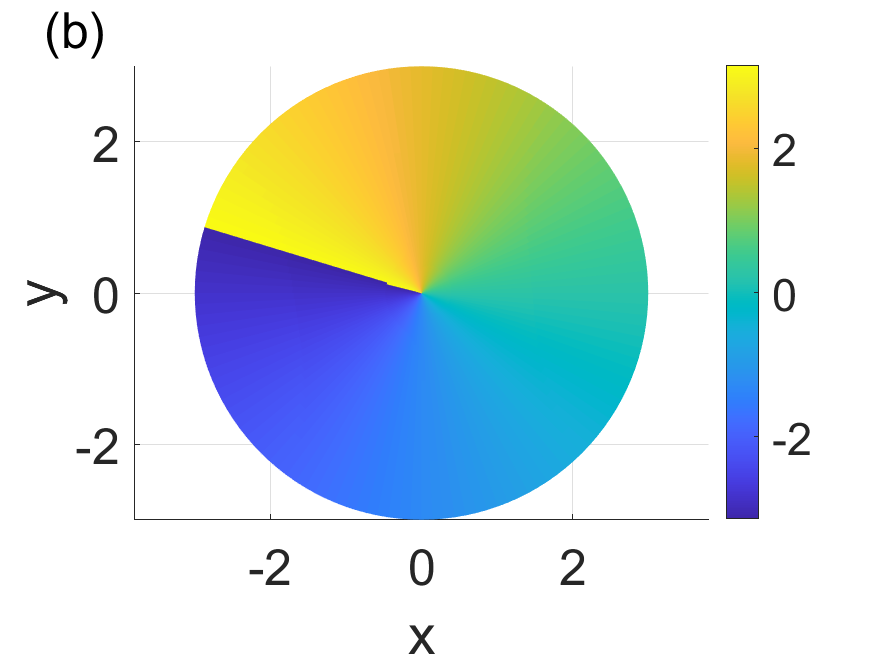}
\includegraphics[width=4.1cm,height=3.27cm,angle=-0]{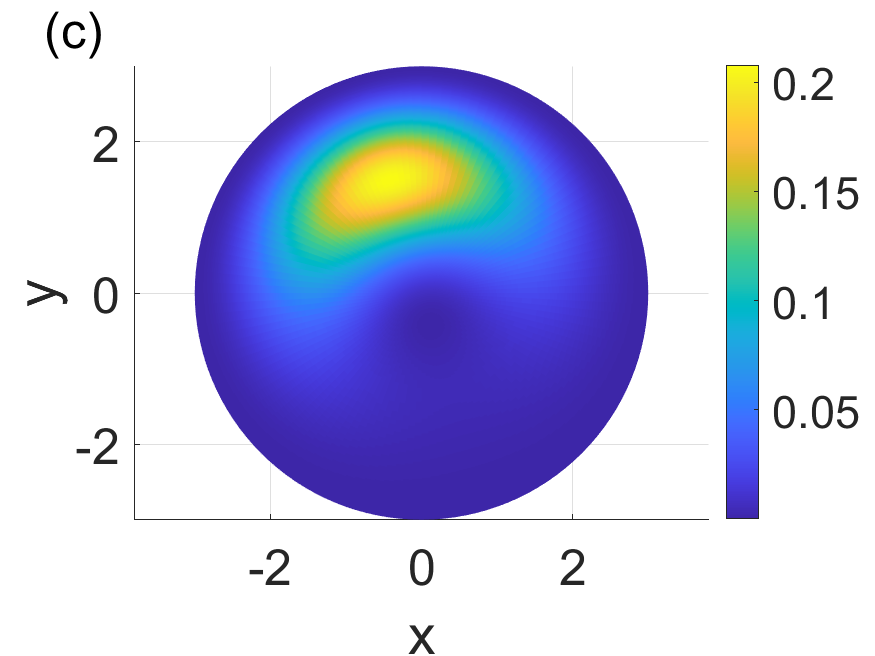}
\includegraphics[width=4.1cm,height=3.27cm,angle=-0]{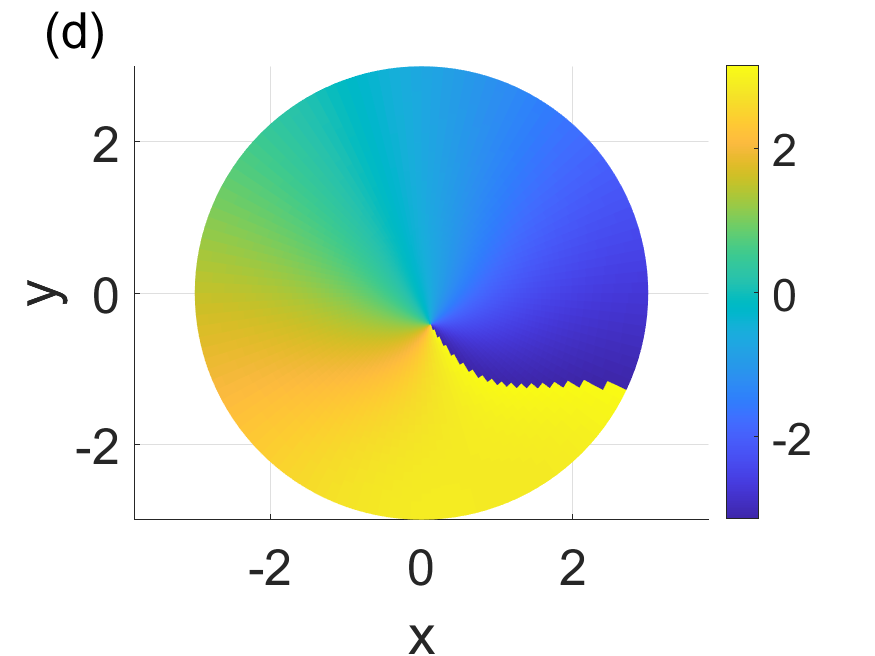}
\caption{(Colour online). (a), (c) Density and (b), (d) phase of the order parameter, 
in a hard-wall potential, with $R/R_0=3$, $\Omega/\Omega_0 = 0.6$, and (a), (b) $g=-0.01$; 
(c), (d) $g=-0.1$.  Here the axes are measured in units of $R_0$, and the density in units 
of $R_0^{-2}$.}
\end{figure}

\begin{figure}
\includegraphics[width=4.1cm,height=3.27cm,angle=-0]{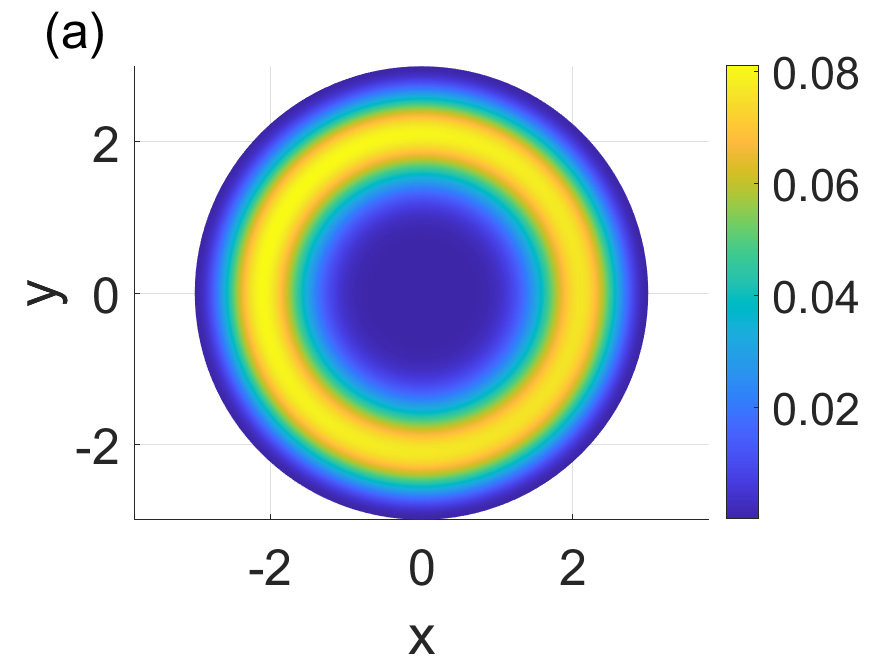}
\includegraphics[width=4.1cm,height=3.27cm,angle=-0]{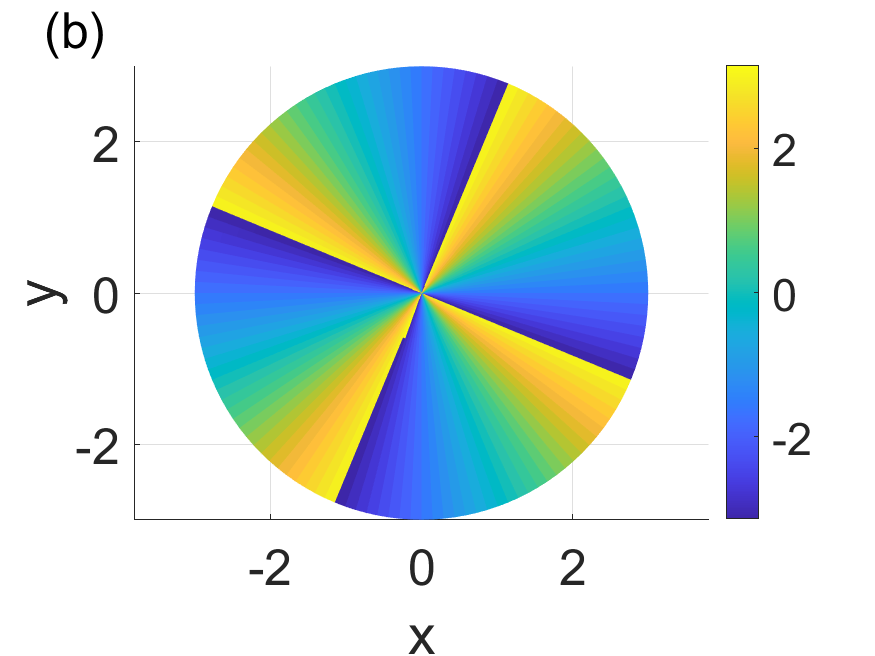}
\includegraphics[width=4.1cm,height=3.27cm,angle=-0]{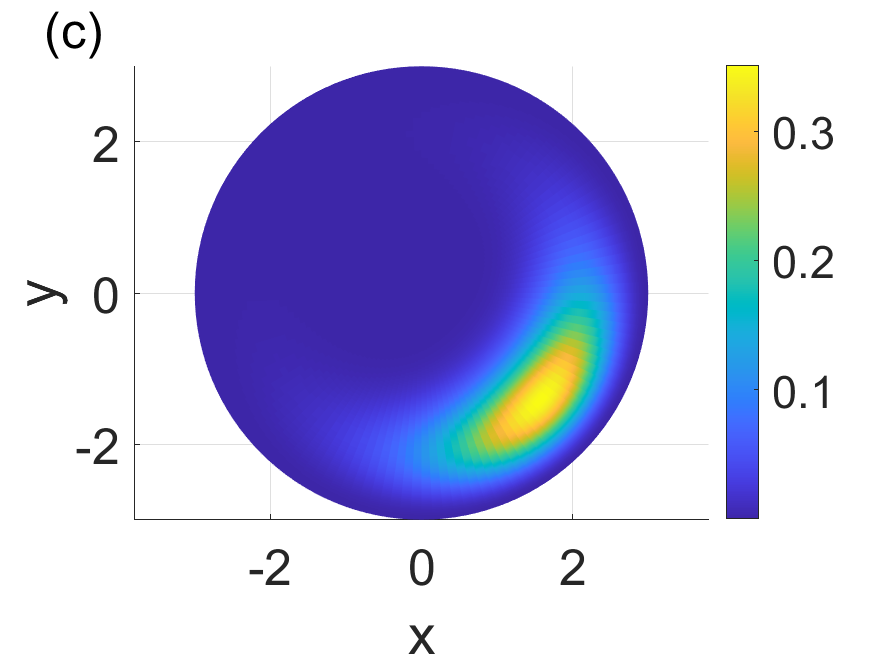}
\includegraphics[width=4.1cm,height=3.27cm,angle=-0]{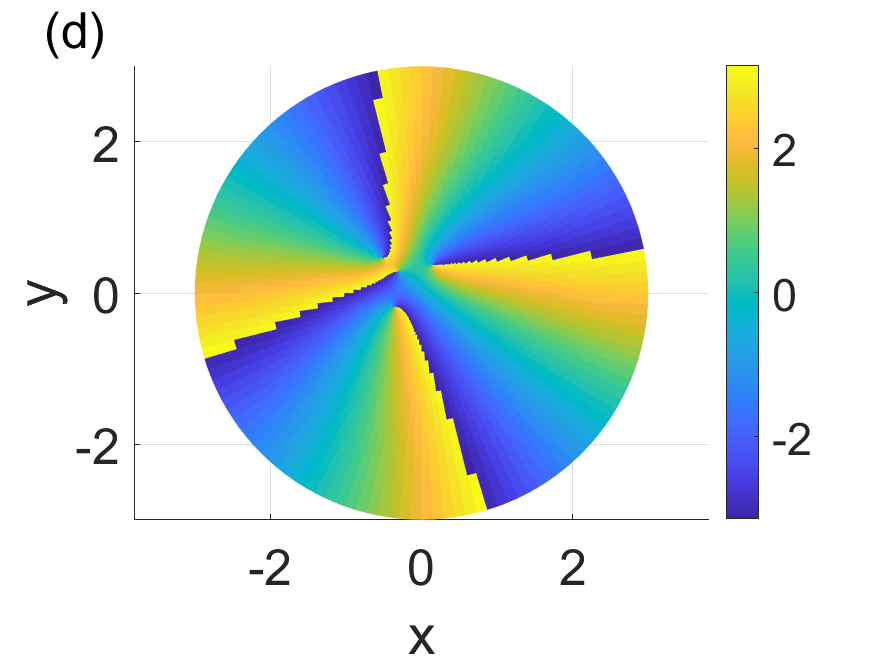}
\caption{(Colour online). (a), (c) Density and (b), (d) phase of the 
order parameter, in a hard-wall potential, with $R/R_0=3$, $\Omega/\Omega_0 = 1.0$, 
and (a), (b) $g=-0.01$; (c), (d) $g=-0.1$.  Here the axes are measured in units 
of $R_0$, and the density in units of $R_0^{-2}$.}
\end{figure}

\section{Experimental relevance} 

The phases that we identified in the present study should be experimentally accessible.
First of all, the assumption of quasi-two-dimensional motion requires that $N a_{\rm sc}
a_z/R^2 \ll 1$, or $a_z \ll R^2/(N a_{\rm sc})$. For a typical box trap of radius $R = 10$ 
$\mu$m, an atom number $N = 5 \times 10^4$ and $a_{\rm sc} = 5$ nm (as in $^{87}\text{Rb}$), 
$a_z$ should be much less than $\approx 0.4$ $\mu$m. For $M$ equal to 87 proton masses, this
translates into $\omega_z \gg 2 \pi \times 0.7$ kHz.

Turning to the coupling constant $g$, this may be tuned, either with use of the Feshbach 
resonances, which will change the scattering length $a_{\rm sc}$ (here $a_{\rm sc}$ may 
also change sign), by tuning the atom number $N$, or by tuning $a_z$. 

For an atom number $N = 5 \times 10^4$, a value of $a_z = 0.4$ $\mu$m, and a scattering length 
$a_{\rm sc} \approx 5$ nm, $g = (N a_{\rm sc}/a_z)/\sqrt{2 \pi}$ is $\approx 250$ (upper part of 
the phase diagram of Fig. 5). Reducing $N$ by one order of magnitude, i.e., to $5 \times 10^3$, 
lowers $g$ to $\approx 25$ (lower part of the phase diagram of Fig. 5).

The unit of rotational frequency in our dimensionless units is set by the characteristic 
scale $\Omega_0 = \hbar/(M R_0^2)$. Again for a typical alkali atom such as $^{87}\text{Rb}$ 
confined in a disk of radius $R = 10\ \mu\text{m}$ (i.e., $R_0 \approx 3.3\ \mu\text{m}$), 
this scale corresponds to $\approx 2 \pi \times 10$ Hz. Rotation can be induced experimentally 
either by phase-imprinting, see, e.g., \cite{pi1, pi2, pi3}, by applying an asymmetric stirring 
beam, or by rotating a weak perturbation at the hard-wall boundary.

\section{Summary and conclusions}

In this work we investigated the rotational properties and the global phase 
diagram of a quasi-two-dimensional Bose-Einstein condensate confined in a hard-wall 
trapping potential. We chose to tune two parameters, namely the rotational 
frequency of the potential $\Omega$ and the dimensionless coupling constant $g$. 
Unlike standard harmonic confinement, see, e.g., Ref.\,\cite{Fetter2009}, the 
hard-wall potential leads to a single-particle energy spectrum that has a positive 
curvature as a function of the single-particle angular momentum. In the rotating 
frame, this positive curvature gives rise to a unique minimum in the single-particle 
energy, stabilizing multiply-quantized vortex states for zero or weak (repulsive
and attractive) interatomic interactions. 

By systematically examining the parameter space of $\Omega$ and $g$, we  
identified seven distinct physical phases across the repulsive and attractive 
interaction regimes. For zero or weak interactions, increasing $\Omega$ induces 
discontinuous phase transitions between multiply-quantized vortex states with 
increasing circulation. As the coupling $g$ increases at fixed $\Omega$, these 
multiply-quantized states undergo a continuous transition into a mixed phase, 
which is characterized by a linear superposition of three single-particle orbitals. 
This mixed state always involves the single-particle state with $m=0$ and as a 
result the density is nonzero at the trap center \cite{e11}. This result, however, 
is specific to the weak-interaction regime.

In the Thomas-Fermi regime the condensate forms a regular vortex lattice. Under 
faster rotation, centrifugal forces deplete the density at the trap center, 
producing a vortex lattice with a central hole. Eventually the condensate forms 
a giant vortex at even larger rotational speeds. 

For effective attractive interactions the hard-wall potential stabilizes the gas 
against collapse for weak coupling, where the system remains in a multiply-quantized 
vortex state, which is dominated by the single-particle kinetic energy. With increasing 
$|g|$, it becomes energetically favourable for the cloud to break the axial symmetry of
the Hamiltonian and form a localized blob, corresponding to a center-of-mass-like state. 
Finally, for sufficiently strong attractive interactions the system collapses.  

The present results also provide a connection with previous studies of rotating 
condensates in different interaction regimes. For strong repulsive interactions, 
Fischer and Baym \cite{e4} studied a rapidly-rotating condensate in a hard-wall 
bucket within the Thomas-Fermi approximation and predicted the sequence from a 
vortex lattice to a vortex lattice with a central hole and, eventually, to a 
giant-vortex state. Our numerical results recover this sequence and connect it 
continuously to the weak- and intermediate-coupling regimes, where 
multiply-quantized vortex and mixed states are present. On the attractive side, 
weak attractive interactions support multiply-quantized vortex states, whereas 
stronger attraction favours a localized, center-of-mass-like state, followed 
eventually by collapse. 

The resulting phase diagram is, to some extent, similar to that found previously 
for a condensate in a quadratic-plus-quartic potential \cite{KJB}. In this case, 
the mixed phase is present also for attractive interactions, whereas in the 
present case this phase is absent, as we saw above. Thus, the present hard-wall 
phase diagram provides a unified framework that connects several limiting regimes 
identified previously in different interaction and confinement regimes.

It is important to mention that, within the present quasi-two-dimensional model 
with hard-wall confinement, the phase diagrams of Figs.\,1, 5, and 8 are universal 
when expressed in terms of the two dimensionless parameters $g$ and $\Omega R^2$. 
This follows directly from the energy functional of Eq.\,(\ref{functional}): upon 
rescaling the spatial coordinates by the radius $R$, the kinetic and interaction 
energies both scale as $R^{-2}$ \cite{PR}, while the two other terms are invariant 
\cite{footnote}. Therefore, the phase diagrams shown in Figs.\,1, 5, and 8 (for 
$R/R_0=3$) determine the corresponding phase diagrams for any value of $R/R_0$, 
simply by rescaling the horizontal axis according to the value of $\Omega R^2$.

Our results demonstrate how a hard-wall potential affects the rotational properties
of a Bose-Einstein condensate, giving rise to a rather rich phase diagram. Given 
the recent advances in the realization and control of box traps, the predicted phases 
should be accessible to experimental investigation.


\begin{thebibliography}{55}
\bibitem{footnote} This scaling is also evident in Eq.\,(\ref{erottot2}), 
which is valid in the weak-coupling limit, as well as in Eqs.\,(\ref{gphb}), 
(\ref{ggg}), and (\ref{gggg}) in the strong-coupling limit.
\end{thebibliography}
\end{document}